\documentclass[prd, 12pt, superscriptaddress, nofootinbib,preprintnumbers,reprint, onecolumn,floatfix]{revtex4-2} 
\usepackage{amsmath,amssymb}
\usepackage{mathtools}
\usepackage{float}
\usepackage{graphicx}
\usepackage{dcolumn}
\usepackage{bm}
\usepackage{natbib}
\usepackage{url}
\usepackage{tikz}
\usetikzlibrary{shapes}
\usetikzlibrary{positioning}
\usetikzlibrary{calc} 
\usepackage{qcircuit}
\usepackage{braket}
\usepackage{bbm}
\usepackage{esvect}
\usepackage{ulem}
\usepackage[colorlinks=true,backref=false, linktocpage=true,
citecolor=blue,urlcolor=blue,linkcolor=blue,pdfpagemode=UseOutlines]{hyperref}
\usepackage{multirow}
\usepackage[section]{placeins}
\usepackage{booktabs}
\usepackage{appendix}
\usepackage{empheq}
\usepackage{array}
\usepackage{subfig}
\usepackage{subcaption}
\usepackage{caption}
\usepackage{todonotes}
\usepackage{amsthm}
\usepackage{slashed}
\usepackage{comment}
\usepackage{makecell}
\usepackage{microtype} 

\tikzset{
  thickdot/.style={circle, fill=black, minimum size=6pt, inner sep=0pt, draw},
  connectline/.style={line width=0.8pt}
}

\begin{document}


\title{Quantum Simulation of the Real-time Dynamics in the multi-flavor Gross-Neveu Model at the utility scale using Superconducting Quantum Computers}
 
\newcommand*{\KU}{Department of Physics and Astronomy, University of Kansas, Lawrence, Kansas 66045, USA.}\affiliation{\KU}
\newcommand*{\Yonsei}{Department of Physics, Yonsei University, 50 Yonsei-ro, Seodaemun-gu, Seoul 03722, South Korea.}\affiliation{\Yonsei}
\newcommand*{\CSIBNL}{Computational Science Department, Brookhaven National Laboratory, Upton, New York 11973, USA.}\affiliation{\CSIBNL}

\author{Talal Ahmed Chowdhury}\email[Corresponding author: ]{talal@ku.edu}\affiliation{\KU}
\author{Seokwon Choi}\email{seokwon0106@yonsei.ac.kr}\affiliation{\Yonsei}
\author{Kyoungchul Kong}\email{kckong@ku.edu}\affiliation{\KU}
\author{Kwangmin Yu}\email[Corresponding author: ]{kyu@bnl.gov}\affiliation{\CSIBNL}

\begin{abstract}
We present a scalable quantum simulation framework for real-time dynamics of the multi-flavor Gross–Neveu model in 1+1 dimensions. Using superconducting quantum processors at utility scale, we develop a hardware-efficient Trotterization whose per-step circuit depth scales with fermion flavor number rather than total system size, enabling simulations beyond 100 qubits. A central contribution of this work is the Localized Diagonal Operator Approximation (LDOA), which systematically reduces the overhead associated with quartic interactions. We formulate diagonal unitary synthesis as a structured least-squares problem in phase space and obtain analytic solutions via the Moore–Penrose pseudoinverse. This formulation provides a principled and quantitatively controlled approximation: in the small Trotter-step regime, the unitary error is directly linked to the phase reconstruction error and vanishes asymptotically as the Trotter step size decreases. This establishes a clear mathematical foundation for the LDOA while significantly reducing two-qubit gate counts and circuit depth, and is broadly applicable to diagonal quantum operators with long-range structure, making it particularly well suited for quantum hardware with limited qubit connectivity. Using these techniques, we run large-scale simulations on IBM superconducting processors and study real-time observables, including density-density correlators. We benchmark against exact diagonalization and tensor network-based methods, finding strong agreement across system sizes. These results show that combining hardware-aware circuit design with rigorous approximations enables practical near-term simulation of interacting fermionic field theories and provides a scalable pathway toward more complex quantum field theory simulations.
\end{abstract}

\maketitle

\section{Introduction}\label{sec:introduction}
Quantum computers are expected to play a pivotal role in investigating the intricate dynamics of quantum field theories. As such, it is vital to validate existing findings on current quantum computers to better understand their potential for future applications. The emergence of real quantum devices as effective, programmable platforms presents an exciting opportunity to precisely prepare complex quantum systems. These platforms serve as invaluable testing grounds, enabling researchers to explore novel phenomena that would be difficult, if not impossible, to observe using traditional experimental     techniques~\cite{Swan-dynamics-quantum-info-review, Georgescu:2013oza, Daley:2022eja}. Therefore, conducting quantum simulations of quantum field theories~\cite {Byrnes-Yamamoto-lattice, Jordan-Lee-Preskill-science, Jordan-Lee-Preskill-scalar, Jordan-Lee-Preskill-fermion, Jordan-Krovi-Lee-Preskill-BQP} on a quantum computer can be a worthwhile objective.

The Gross-Neveu model is a $1+1$-dimensional field theory of Dirac fermions with $N$ flavors, interacting via quartic interactions. It shares several fundamental characteristics with Quantum Chromodynamics (QCD), including renormalizability, asymptotic freedom, dynamical symmetry breaking of a $Z_{2}$ chiral symmetry, and dynamical mass generation~\cite{Gross-Neveu}. Notably, the Gross-Neveu model also exhibits a rich spectrum of particles that can be interpreted as bound states of fermions and antifermions, as well as multifermion bound states~\cite{Dashen-Hasslacher-Neveu}. Additionally, the scattering matrix (S-matrix) in the Gross-Neveu model, which relates the initial and final multiparticle states involved in scattering processes, can be completely factorized into two-particle S-matrices, which arise from the existence of an infinite number of conservation laws~\cite{Zamolodchikov-PLB, Zamolodchikov:1978xm, Witten:1977xv}. Furthermore, in the limit of large $N$, the Gross-Neveu model reveals three distinct phases at finite temperature $T$ and chemical potential $\mu$: a massive phase, where a homogeneous chiral condensate breaks the chiral symmetry (characterized by low $T$ and low $\mu$); a massless symmetric phase with a vanishing condensate (found at high $T$ or high $\mu$); and an inhomogeneous phase, also known as the kink-antikink crystal phase, where the condensate oscillates in the spatial direction (observed at low $T$ and high $\mu$)~\cite{Thies-Urlichs-revised-phase, Schnetz-Thies-Urlichs, Lenz-Pannullo-Bjorn-Wipf, Lenz-Pannullo-Wagner-Bjorn-Wipf-baryon}. Moreover, the Gross-Neveu model effectively represents cold dense nuclear matter~\cite{Lajer-Konik-Pisarski-Tsevlik}. Given the model's significant importance and its connection to certain regimes of QCD, we aim to focus on the quantum simulation of real-time dynamics within this model, capturing its essential features during time evolution, using IBM's superconducting quantum computers.

Simulating the real-time dynamics of strongly interacting fermionic theories, such as the Gross-Neveu model, using Monte Carlo methods encounters significant limitations in time evolution computations due to the onset of the dynamical sign problem, as highlighted in Ref.~\cite{Ortiz-Gubernatis-Knill-Laflamme}. In contrast, quantum simulations using quantum computers are free from these issues. Some studies have explored the quantum simulation of the Gross-Neveu model on real quantum computers with small system sizes~\cite{QS-Gross-Neveu}. In contrast, we have developed a robust quantum simulation framework that implements time evolution in Minkowski spacetime and features scalable, optimized Trotterization circuits for quantum computers, particularly those with limited qubit connectivity. 
A central component of this framework is the Localized Diagonal Operator Approximation (LDOA), which provides a systematic and hardware-efficient method for approximating quartic interaction terms in fermionic Hamiltonians. By reformulating diagonal unitary synthesis as a structured optimization problem in phase space, LDOA enables controlled circuit compression while preserving asymptotic consistency in the small Trotter-step limit. This formulation is not limited to the Gross–Neveu model but applies broadly to diagonal quantum operators with long-range interactions, independent of the underlying physical model, and is particularly well suited to implementations on quantum hardware with limited connectivity.
Importantly, the circuit depth remains constant even as the total number of qubits increases. As a result, our quantum simulation enables us to accurately compute the real-time dynamics of physically relevant observables, such as the density-density correlation functions in the Gross-Neveu model for systems involving over a hundred qubits, hence mitigating the finite-size effects that are prevalent in smaller systems. Additionally, we probe the dynamics of entanglement entropy in the Gross-Neveu model by measuring the second R\'enyi entropy over time on IBM quantum computers, thereby unveiling the nature of quantum information scrambling within this framework.

The manuscript is organized as follows. In section~\ref{sec:Gross-Neveu}, we present a concise introduction to the Gross-Neveu model, its lattice discretization, and mapping onto qubits. Section~\ref{sec:Trotterization} presents the Trotterization circuits for the time evolution in the lattice Gross-Neveu model. In section~\ref{sec:DOA}, we present the LDOA framework, providing the hardware-efficient method for approximating the quartic interaction term of the Gross-Neveu model. Section~\ref{sec:IBM-Trotterization} provides the details of the implementation of the Trotter circuits in IBM quantum computers. In section~\ref{sec:Experimental-Results}, we present the variation of the density-density correlations in the lattice Gross-Neveu model for system size up to more than a hundred qubits under the time evolution using our LDOA-implemented Trotterization circuits and their validation using classical exact and Tensor network-based methods. Section~\ref{sec:EEM_results} presents measuring the entanglement entropy dynamics in the Gross-Neveu model using IBM quantum computers. Finally, we conclude in section~\ref{sec:conclusion}. In appendix~\ref{app:error_mitigations}, we provide a detailed discussion of the quantum error mitigation methods used in our study.

\section{The Gross-Neveu Model}\label{sec:Gross-Neveu}
The Gross-Neveu model~\cite{Gross-Neveu} is described by the Lagrangian,
\begin{equation}
    \mathcal{L}=\sum_{b=0}^{N-1} i\overline{\psi}_{b}\gamma^{\mu}\partial_{\mu}\psi_{b}+\frac{g^{2}}{2}\left(\sum_{b=0}^{N-1}\overline{\psi}_{b}\psi_{b}\right)^{2}\,,\label{eq:GN-Lagrangian}
\end{equation}
where $N$ is the number of fermion flavors, $\psi_{b}$ is the massless Dirac fermion of flavor $b$, and the quartic interaction is controlled by the coupling $g$. The two-component Dirac field of flavor $b$ is
\begin{equation}
    \psi_{b}(x)=\begin{pmatrix}
        \psi_{1,b}(x)\\
        \psi_{2,b}(x)
    \end{pmatrix}
\end{equation}
where $\psi_{1,b}(x)$ and $\psi_{2,b}(x)$ are Grassmann fields. The $\gamma$ matrices follow $\{\gamma^{\mu},\,\gamma^{\nu}\}=2\eta^{\mu\nu}I_{2}$, where $\eta^{\mu\nu}=\text{diag}(1,-1)$ is the Minkowski metric in 1+1D, and in the Dirac representation they are given as $\gamma^{0}=Z$, $\gamma^{1}=iY$, $\gamma^{5}=\gamma^{0}\gamma^{1}=X$ with $X,Y,Z$ being the Pauli matrices. Moreover, the Lagrangian is invariant under the discrete $\gamma_{5}$ transformation,
\begin{equation}
    \psi_{b}\rightarrow \gamma_{5}\psi_{b}\,,\label{eq:discrete-chiral}
\end{equation}
which ensures the masslessness of the fermion field to any order of perturbation theory.

The Hamiltonian of the Gross-Neveu model is given by
\begin{equation}
    H = \int dx \left[-i \sum_{b=0}^{N-1}\bar{\psi}_{b}\gamma^{1}\partial_{x}\psi_{b}-\frac{g^2}{2}\left(\sum_{b=0}^{N-1}\bar{\psi}_{b}\psi_{b}\right)^{2}\right]\,.\label{eq:GN-Hamiltonian}
\end{equation}

Quantum field theories, such as the Gross-Neveu model, have an infinite number of degrees of freedom, and handling them computationally requires spatial lattice discretization. Therefore, in this work, we use the Kogut-Susskind lattice fermion prescription~\cite{Kogut-Susskind, Banks-Susskind-Kogut, susskind-lattice-fermion} to express the fermion on the lattice points. At first, we discretize the spatial line with $L_D$ points and lattice spacing $a$. Now the Dirac fermion at lattice point $x=n a$, where $a$ is the lattice spacing with dimensions of length, is $\psi(x)\equiv \psi(x=na)$ whose components are 
\begin{equation}
\psi_{b}(x= n a)=\begin{pmatrix}
    \psi_{1,b}(na)\\
    \psi_{2,b}(na)
\end{pmatrix}
\end{equation}

Now following the Kogut-Susskind lattice fermion prescription, we map the upper and lower fermion components of the Dirac fermion at $x = n a$ onto the staggered sites $(2n, 2n+1)$ (for a similar treatment, see Ref.~\cite{Roose-et-al-lattice-GN}), respectively, as follows
\begin{align}
    \psi_{1,b}(na)&\mapsto \frac{1}{\sqrt{a}}\phi_{2n,b}\,,\label{eq:staggered-upper}\\
    \psi_{2,b}(na)&\mapsto \frac{1}{\sqrt{a}}\phi_{2n+1,b}\,,\label{eq:staggered-lower}
\end{align}    
where, we place the upper component $\psi_{1,b}(na)$ to the site with index $(2n,b)$ and the lower component $\psi_{2,b}(na)$ to the site with index $(2n+1,b)$. Here, $\phi_{j,b}$ and $\phi^{\dagger}_{j,b}$ are the fermionic annihilation and creation operators (dimensionless) at a site $j$ and flavor $b$. Therefore, the upper and lower components of Dirac fermions on $L_{D}$ points are mapped into $L=2L_{D}$ lattice sites. Furthermore, the spatial derivatives are expressed as
\begin{align}
    \partial_{x}\psi_{1,b}(na)&\mapsto \frac{1}{2a^{3/2}}(\phi_{2n+2,b}-\phi_{2n,b})\,\,\,(\mathrm{forward\,\,\,derivative})\,,\label{eq:forward}\\
    \partial_{x}\psi_{2,b}(na)&\mapsto \frac{1}{2a^{3/2}}(\phi_{2n+1,b}-\phi_{2n-1,b})\,\,\,(\mathrm{backward\,\,\,derivative})\,.\label{eq:backward}
\end{align}
Furthermore, the fermionic creation and annihilation operators follow the anticommutation relations,
\begin{equation}
\{\phi_{j,b},\phi_{j',b'}\}=\{\phi^{\dagger}_{j,b},\phi^{\dagger}_{j',b'}\}=0,\,\,\,\{\phi_{j,b},\phi^{\dagger}_{j',b'}\}=\delta_{j j'}\delta_{bb'}\,.\label{eq:anticomm}
\end{equation}

Now the quadratic Hamiltonian containing the spatial derivative is,
\begin{equation}
    H_{\text{quadratic}}= -i\sum_{b}a\sum_{n=0}^{L_D-1} \left(\psi^{\dagger}_{1,b}(na)\partial_x\psi_{2,b}(na)+\psi^{\dagger}_{2,b}(na)\partial_{x}\psi_{1,b}(na)\right)\,.\label{eq:quadratic} 
\end{equation}
Using Eq.~(\ref{eq:forward}) and Eq~(\ref{eq:backward}), in conjunction with boundary conditions, $\psi_{1,b}(na)=\psi_{2,b}(na)=0$ for $n<0$ and $n>L_{D}-1$, we have
\begin{equation}
    H_{\text{quadratic}}=-\frac{i}{2a}\sum_{j=0}^{2L_{D}-2}\sum_{b=0}^{N-1}\left(\phi^{\dagger}_{j,b}\phi_{j+1,b}-\phi^{\dagger}_{j+1,b}\phi_{j,b}\right)\,.\label{eq:Hopping}
\end{equation}
Now the bilinear term $\overline{\psi}_{b}\psi_{b}$ in terms of upper and lower component fields is
\begin{equation}
\overline{\psi}_{b}\psi_{b}=\psi^{\dagger}_{1,b}\psi_{1,b}-\psi^{\dagger}_{2,b}\psi_{2,b}\,.\label{eq:bilinear}
\end{equation}
Using Eq.~(\ref{eq:staggered-upper}) and Eq.~(\ref{eq:staggered-lower}) and defining the fermionic number density operator at the lattice site $j$ for a specific flavor $b$ as $\rho_{j,b}=\phi^{\dagger}_{j,a}\phi_{j,a}$, the quartic interaction term of the Hamiltonian becomes,
\begin{align}
    H_{\mathrm{quartic}}&=-\frac{g^{2}}{2a}\sum_{n=0}^{L_{D}-1}\sum_{b=0}^{N-1}\left(\rho_{2n,b}+\rho_{2n+1,b}-2\rho_{2n,b}\rho_{2n+1,b}\right)\nonumber\\
    &-\frac{g^{2}}{a}\sum_{n=0}^{L_{D}-1}\sum_{\substack{b,c=0,\\b<c}}^{N-1}\left(\rho_{2n,b}\rho_{2n,c}+\rho_{2n+1,b}\rho_{2n+1,c}-\rho_{2n,b}\rho_{2n+1,c}-\rho_{2n+1,b}\rho_{2n,c}\right)\label{eq:quartic}
\end{align}
For a single lattice site $j$ where $j=0,1,...,L$ with $L=2L_{D}$, and $N$ flavors we can have the following states,
\begin{eqnarray}
    |0\rangle,\,\,\, \phi^{\dagger}_{j,b}|0\rangle,\,\,\,\,\phi^{\dagger}_{j,b}\phi^{\dagger}_{j,c}|0\rangle\,\,\,(b<c),\,\,\,\,\phi^{\dagger}_{j,b}\phi^{\dagger}_{j,c}\phi^{\dagger}_{j,d}|0\rangle\,\,\,(b<c<d),\,\,\,.....\,\,\,\phi^{\dagger}_{j,0}\phi^{\dagger}_{j,1}...\phi^{\dagger}_{j,N-1}|0\rangle\,.\label{eq:single-site-Hilbert}
\end{eqnarray}
Then the dimension of the space of states $\cal{H}$ associated with the single site $j$ (which is same for all $j$) is
\begin{equation}
\text{dim}\mathcal{H}=\sum_{k=0}^{N} {}^{N}C_{k}\,,\label{eq:dim-hj}
\end{equation}
Therefore, the total dimension of the Hilbert space associated with $L$ fermionic lattice sites is $(\text{dim}\mathcal{H})^{L}$.

We use the Jordan-Wigner transformation~\cite{jordan-wigner} to map the fermionic creation and annihilation operators onto qubit operators as follows,
\begin{eqnarray}
    \phi_{j,b}&\mapsto & \frac{1}{2}Z_{0,b}\otimes Z_{1,b}\otimes...\otimes Z_{j-1,b}\otimes(X_{j,b}+iY_{j,b})\,,\nonumber\\
    \phi^{\dagger}_{j,b}&\mapsto & \frac{1}{2}Z_{0,b}\otimes Z_{1,b}\otimes...\otimes Z_{j-1,b}\otimes (X_{j,b}-iY_{j,b})\,.\label{eq:JW-map}
\end{eqnarray}

Moreover, the fermionic site with index $(j,b)$ is mapped to qubit $q_{j,b}$ as
\begin{equation}
    q_{j,b} \rightarrow q_{jN+b}
    \label{eq:qubit-mapping}
\end{equation}
where $N$ is the total flavors. Hence, we require total $N_{\text{qubits}}=N\,L$ number of qubits to describe the lattice Gross-Neveu model with $L$ spatial lattice sites and $N$ flavors. Specifically, in our qubit mapping the staggered fermionic sites $(2n, 2n+1)$ associated with the Dirac fermion at site $n$ are concatenated by the flavor degrees of freedom as $[(2n,0), (2n,1),...,(2n,N-1), (2n+1, 0), (2n+1,1),...,(2n+1, N-1)]$ where each $(j,b)$ represents the fermionic site at $j$ with flavor $b$. As an example, for $N=3$ case, the qubit assignment for a single site $j$ is given as,
\begin{align}
    |0\rangle&\mapsto |000\rangle,\,\,c^{\dagger}_{j,0}|0\rangle\mapsto |100\rangle,\,\,c^{\dagger}_{j,1}|0\rangle\mapsto |010\rangle,\,\,c^{\dagger}_{j,2}|0\rangle\mapsto |001\rangle\,\nonumber\\
    c^{\dagger}_{j,0}c^{\dagger}_{j,1}|0\rangle&\mapsto |110\rangle,\,\,c^{\dagger}_{j,0}c^{\dagger}_{j,2}|0\rangle\mapsto |101\rangle,\,\,c^{\dagger}_{j,1}c^{\dagger}_{j,2}|0\rangle\mapsto |011\rangle,\,\,c^{\dagger}_{j,0}c^{\dagger}_{j,1}c^{\dagger}_{j,2}|0\rangle\mapsto |111\rangle\,.\label{eq:three-flavor}
    \end{align}

The JW-transformed quadratic term of the lattice Gross-Neveu Hamiltonian is,
\begin{align}
        \tilde{H}_{\text{quadratic}} &= \sum_{j=0}^{L-2} \sum_{b=0}^{N-1}\frac{1}{4a}(X_{j,b}Y_{j+1,b}-Y_{j,b}X_{j+1,b})\,.\label{eq:JW-quadratic}
\end{align}
Now using $\rho_{j,b}\mapsto \frac{1}{2}(I-Z)_{j,b}\equiv P^{-}_{j,b}$, we have JW-transformed quartic term of the lattice Gross-Neveu Hamiltonian from Eq.~(\ref{eq:quartic}) as
\begin{align}
    \tilde{H}_{\text{quartic}}&=-\frac{g^{2}}{2a}\sum_{n=0}^{L_{D}-1}\sum_{b=0}^{N-1}(P^{-}_{2n,b}+P^{-}_{2n+1,b}-2P^{-}_{2n,b}P^{-}_{2n+1,b})\nonumber\\
    &-\frac{g^{2}}{a}\sum_{n=0}^{L_{D}-1}\sum_{\substack{b,c=0,\\b<c}}(P^{-}_{2n,b}P^{-}_{2n,c}+P^{-}_{2n+1,b}P^{-}_{2n+1,c}-P^{-}_{2n,b}P^{-}_{2n+1,c}-P^{-}_{2n+1,b}P^{-}_{2n,c})\,,\label{eq:JW-quartic-0}
\end{align}
which can be further simplified into 
\begin{equation}
    \tilde{H}_{\text{quartic}}=\frac{g^{2}}{4a}\sum_{n=0}^{L_{D}-1}\left[\sum_{b=0}^{N-1}Z_{2n,b}Z_{2n+1,b}-\sum_{\substack{b,c=0,\\ b<c}}^{N-1}\left(Z_{2n,b}Z_{2n,c}+Z_{2n+1,b}Z_{2n+1,c}-Z_{2n,b}Z_{2n+1,c}-Z_{2n+1,b}Z_{2n,c}\right)\right]-\frac{g^{2}}{4a}L_{D}N\,,\label{eq:JW-quartic}
\end{equation}
Consequently, we have the JW-transformed lattice Gross-Neveu Hamiltonian acting on $N_{\text{qubits}}=L\,N$ qubits,
\begin{equation}
\tilde{H}=\tilde{H}_{\text{quadratic}}+\tilde{H}_{\text{quartic}}\,.\label{eq:JW-GN-Hamiltonian}
\end{equation}
Moreover, by using the following unitary transformation (see, for example Refs.~\cite{Oshikawa-Affleck, derzhko2004effects}),
\begin{equation}
\hat{X}_{j,b}=X_{j,b}\cos\phi_{j}+Y_{j,b}\sin\phi_{j},\,\,\,\hat{Y}_{j,b}=-X_{j,b}\sin\phi_{j}+Y_{j,b}\cos\phi_{j},\,\,\,\hat{Z}_{j,b}=Z_{j,b}\,,\label{eq:XY-unitary}
    \end{equation}
with $\phi_{j}=j\pi/2$, we have
\begin{equation}
        \tilde{H}'_{\text{quadratic}} = \sum_{j=0}^{L-1} \sum_{b=0}^{N-1}\frac{1}{4a}(\hat{X}_{j,b}\hat{X}_{j+1,b}+\hat{Y}_{j,b}\hat{Y}_{j+1,b})\,,\label{eq:JW-quadratic-XXZ-form}
\end{equation}
whereas $\tilde{H}_{\text{quartic}}$ in Eq.~(\ref{eq:JW-quartic-0}) or Eq.~(\ref{eq:JW-quartic}) remains the same. So, we can see that the JW-transformed lattice Gross-Neveu Hamiltonian resembles XXZ spin chain interactions.

In this work, we study the time evolution of density-density correlator,
\begin{equation}
    \hat{C}_{\text{density}}(j,j')=\frac{1}{a^2}\sum_{b=0}^{N-1}(-1)^{j+j'}\rho_{j,b}\rho_{j',b}\,.\label{eq:density-density}
\end{equation}
Using the qubit mapping, we have
\begin{equation}
    \tilde{C}_{\text{density}}(j,j')=\frac{1}{4a^{2}}\sum_{b=0}^{N-1}(-1)^{j+j'}(I-Z)_{j,b}(I-Z)_{j',b}\,.\label{eq:JW-density-density}
\end{equation}
We start with an initial quantum state $|\psi_{0}\rangle$ that is not an eigenstate of the lattice Gross-Neveu Hamiltonian $H$. Under time evolution, this state transforms to $\psi(t)\rangle = e^{-iHt/\hbar}|\psi_{0}\rangle$. We then compute the expectation value of the density-density correlator, given by $C_{\text{density}}(j,j',t) = \langle \psi(t)|\hat{C}_{\text{density}}(j,j')|\psi(t)\rangle$, which is referred to in the remainder of the paper as the density-density correlation, respectively, in order to explore the real-time dynamics of the Gross-Neveu model.

\section{Real-time Dynamics of Gross-Neveu Model via Trotterization}\label{sec:Trotterization}

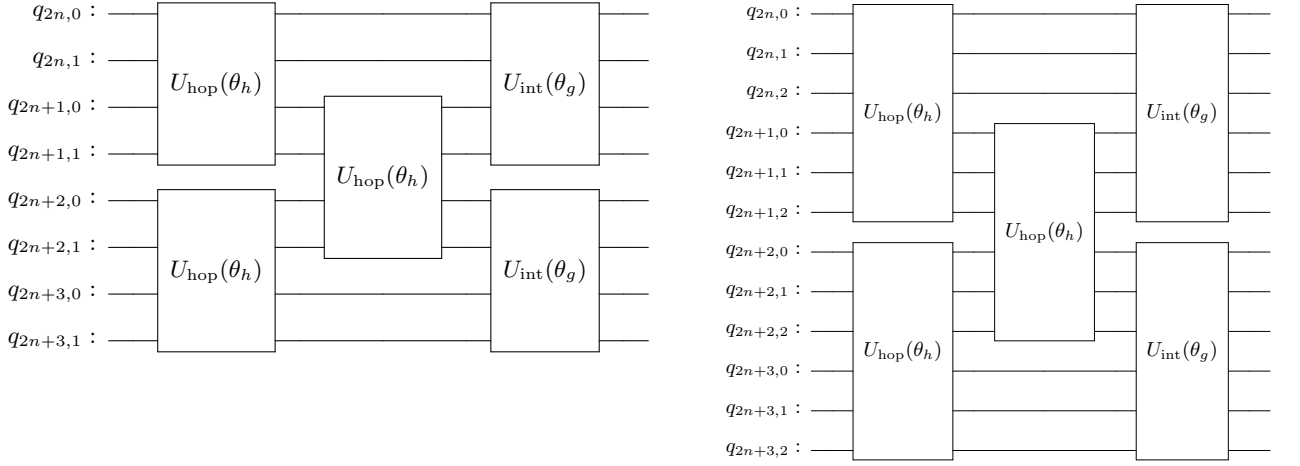
\begin{figure}
    \centerline{
    \scalebox{1.0}{\Qcircuit @C=1.0em @R=1.0em @!R {
    \nghost{q_{2n,0}:} & \lstick{q_{2n,0}:} & \qw & \multigate{3}{U_{\mathrm{hop}}(\theta_{h})} & \qw & \qw & \qw & \multigate{3}{U_{\mathrm{int}}(\theta_{g})} & \qw & \qw \\
    \nghost{q_{2n,1}:} & \lstick{q_{2n,1}:} & \qw & \ghost{U_{\mathrm{hop}}(\theta_{h})}     & \qw & \qw & \qw & \ghost{U_{\mathrm{int}}(\theta_{g})}     & \qw & \qw \\
    \nghost{q_{2n+1,0}:} & \lstick{q_{2n+1,0}:} & \qw & \ghost{U_{\mathrm{hop}}(\theta_{h})}     & \qw & \multigate{3}{U_{\mathrm{hop}}(\theta_{h})} & \qw & \ghost{U_{\mathrm{int}}(\theta_{g})}     & \qw & \qw \\
    \nghost{q_{2n+1,1}:} & \lstick{q_{2n+1,1}:} & \qw & \ghost{U_{\mathrm{hop}}(\theta_{h})}     & \qw & \ghost{U_{\mathrm{hop}}(\theta_{h})}     & \qw & \ghost{U_{\mathrm{int}}(\theta_{g})}     & \qw & \qw \\
    \nghost{q_{2n+2,0}:} & \lstick{q_{2n+2,0}:} & \qw & \multigate{3}{U_{\mathrm{hop}}(\theta_{h})} & \qw & \ghost{U_{\mathrm{hop}}(\theta_{h})}     & \qw & \multigate{3}{U_{\mathrm{int}}(\theta_{g})} & \qw & \qw \\
    \nghost{q_{2n+2,1}:} & \lstick{q_{2n+2,1}:} & \qw & \ghost{U_{\mathrm{hop}}(\theta_{h})}     & \qw & \ghost{U_{\mathrm{hop}}(\theta_{h})}     & \qw & \ghost{U_{\mathrm{int}}(\theta_{g})}     & \qw & \qw \\
    \nghost{q_{2n+3,0}:} & \lstick{q_{2n+3,0}:} & \qw & \ghost{U_{\mathrm{hop}}(\theta_{h})}     & \qw & \qw & \qw & \ghost{U_{\mathrm{int}}(\theta_{g})}     & \qw & \qw \\
    \nghost{q_{2n+3,1}:} & \lstick{q_{2n+3,1}:} & \qw & \ghost{U_{\mathrm{hop}}(\theta_{h})}     & \qw & \qw & \qw & \ghost{U_{\mathrm{int}}(\theta_{g})}     & \qw & \qw 
}}\hspace{5mm}
\scalebox{0.85}{\Qcircuit @C=1.0em @R=1.0em @!R {
    \nghost{q_{2n,0}:}   & \lstick{q_{2n,0}:}   & \qw & \multigate{5}{U_{\mathrm{hop}}(\theta_{h})} & \qw & \qw & \qw & \multigate{5}{U_{\mathrm{int}}(\theta_{g})} & \qw & \qw \\
    \nghost{q_{2n,1}:}   & \lstick{q_{2n,1}:}   & \qw & \ghost{U_{\mathrm{hop}}(\theta_{h})}     & \qw & \qw & \qw & \ghost{U_{\mathrm{int}}(\theta_{g})}     & \qw & \qw \\
    \nghost{q_{2n,2}:}   & \lstick{q_{2n,2}:}   & \qw & \ghost{U_{\mathrm{hop}}(\theta_{h})}     & \qw & \qw & \qw & \ghost{U_{\mathrm{int}}(\theta_{g})}     & \qw & \qw \\
    \nghost{q_{2n+1,0}:} & \lstick{q_{2n+1,0}:} & \qw & \ghost{U_{\mathrm{hop}}(\theta_{h})}     & \qw & \multigate{5}{U_{\mathrm{hop}}(\theta_{h})} & \qw & \ghost{U_{\mathrm{int}}(\theta_{g})}     & \qw & \qw \\
    \nghost{q_{2n+1,1}:} & \lstick{q_{2n+1,1}:} & \qw & \ghost{U_{\mathrm{hop}}(\theta_{h})}     & \qw & \ghost{U_{\mathrm{hop}}(\theta_{h})}     & \qw & \ghost{U_{\mathrm{int}}(\theta_{g})}     & \qw & \qw \\
    \nghost{q_{2n+1,2}:} & \lstick{q_{2n+1,2}:} & \qw & \ghost{U_{\mathrm{hop}}(\theta_{h})}     & \qw & \ghost{U_{\mathrm{hop}}(\theta_{h})}     & \qw & \ghost{U_{\mathrm{int}}(\theta_{g})}     & \qw & \qw \\
    \nghost{q_{2n+2,0}:} & \lstick{q_{2n+2,0}:} & \qw & \multigate{5}{U_{\mathrm{hop}}(\theta_{h})} & \qw & \ghost{U_{\mathrm{hop}}(\theta_{h})}     & \qw & \multigate{5}{U_{\mathrm{int}}(\theta_{g})} & \qw & \qw \\
    \nghost{q_{2n+2,1}:} & \lstick{q_{2n+2,1}:} & \qw & \ghost{U_{\mathrm{hop}}(\theta_{h})}     & \qw & \ghost{U_{\mathrm{hop}}(\theta_{h})}     & \qw & \ghost{U_{\mathrm{int}}(\theta_{g})}     & \qw & \qw \\
    \nghost{q_{2n+2,2}:} & \lstick{q_{2n+2,2}:} & \qw & \ghost{U_{\mathrm{hop}}(\theta_{h})}     & \qw & \ghost{U_{\mathrm{hop}}(\theta_{h})}     & \qw & \ghost{U_{\mathrm{int}}(\theta_{g})}     & \qw & \qw \\
    \nghost{q_{2n+3,0}:} & \lstick{q_{2n+3,0}:} & \qw & \ghost{U_{\mathrm{hop}}(\theta_{h})}     & \qw & \qw & \qw & \ghost{U_{\mathrm{int}}(\theta_{g})}     & \qw & \qw \\
    \nghost{q_{2n+3,1}:} & \lstick{q_{2n+3,1}:} & \qw & \ghost{U_{\mathrm{hop}}(\theta_{h})}     & \qw & \qw & \qw & \ghost{U_{\mathrm{int}}(\theta_{g})}     & \qw & \qw \\
    \nghost{q_{2n+3,2}:} & \lstick{q_{2n+3,2}:} & \qw & \ghost{U_{\mathrm{hop}}(\theta_{h})}     & \qw & \qw & \qw & \ghost{U_{\mathrm{int}}(\theta_{g})}     & \qw & \qw 
}}}
    \caption{The single Trotter-step of the first-order Trotterization circuit highlighting the consecutive fermionic lattice sites, $(2n,\,2n+1,\,2n+2,\,2n+3)$ for $N=2$ flavors (left figure), and $N=3$ flavors (right figure), respectively.}
    \label{fig:single-Trotter-step}
\end{figure}

The time evolution operator generated by the lattice Gross-Neveu Hamiltonian is implemented on a quantum computer using the Trotter-Suzuki decomposition~\cite{trotter, suzuki1, suzuki2}, commonly known as the Trotterization. The first-order Trotterization of the time evolution operator, $U(t)=e^{-i \tilde{H} t/\hbar}$, where $\tilde{H}$ is the Jordan-Wigner-transformed lattice Gross-Neveu Hamiltonian in Eq.~(\ref{eq:JW-GN-Hamiltonian}), acting on $LN$ qubits is given by,
\begin{equation}
	U^{(1)}(t)=\prod_{k=1}^{r}\hat{U}^{(1)}_{\mathrm{Trotter}_{k}}(\delta t),
	\label{eq:first-trotter-1}
\end{equation}
where $\delta t = t/r$ is the Trotter time-step, $r$ is the number of Trotter steps associated with the simulation time $t$, and $\hat{U}^{(1)}_{\mathrm{Trotter}}$ represents a single Trotter step. Now, for the lattice Gross-Neveu Hamiltonian in Eq.~(\ref{eq:JW-GN-Hamiltonian}), denoted as $\tilde{H}=\sum_{s=1}^{M}\tilde{H}_{s}$, where $M$ is the total number of terms in the Hamiltonian, the $\hat{U}^{(1)}_{\mathrm{Trotter}}(\delta t)$ is given as,
\begin{equation}
\hat{U}^{(1)}_{\mathrm{Trotter}}(\delta t)=\prod_{s=1}^{M}e^{-i H_{s}\delta t/\hbar}\,.
\label{eq:single-Trotter-1}
\end{equation}
By using the following operators,
\begin{equation}
    \hat{J}_{p,q} = X_{p}Y_{q}-Y_{p}X_{q},\,\,\,\hat{K}_{p,q}=P^{-}_{p}P^{-}_{q},
    \label{eq:Pauli-ops}
\end{equation}
where, $\hat{J}_{p,q}$ and $\hat{K}_{p,q}$ operators act on $p$ and $q$-th qubits of the qubit register, the unitary operator associated with the quadratic part of the lattice Gross-Neveu Hamiltonian in Eq.~(\ref{eq:JW-GN-Hamiltonian}) can be written as,
\begin{equation}
    e^{-i \frac{\delta t}{4 a \hbar}\sum_{j=0}^{L-2}\sum_{b=0}^{N-1}\hat{J}_{jN+b,(j+1)N+b}}\mapsto \left(\bigotimes_{\substack{j=0\\
    j(\mathrm{mod}\,2)=0}}\left[\bigotimes_{b=0}^{N-1}e^{-i\frac{\theta_{h}}{2}\hat{J}_{jN+b,(j+1)N+b}}\right]\right).\left(\bigotimes_{\substack{j=0\\
    j(\mathrm{mod}\,2)=1}}\left[\bigotimes_{b=0}^{N-1}e^{-i\frac{\theta_{h}}{2}\hat{J}_{jN+b,(j+1)N+b}}\right]\right)\label{eq:quadratic-trotter-decompose}\,,
\end{equation}
where, we denote, $\theta_{h}=\frac{\delta t}{2 a\hbar}$. Besides, $L = 2L_{D}$ is the number of staggered fermionic lattice sites associated with Dirac fermions on $L_D$ lattice sites on the line, and $b=0,1,..,N-1$ denotes the $N$ flavors of the fermion. On the other hand, the unitary operator associated with the quartic interaction terms in Eq.~(\ref{eq:JW-quartic-0}) (or equivalently with those in Eq.~(\ref{eq:JW-quartic})) is written as,
\begin{align}
    &e^{-i \tilde{H}_{\text{quartic}}\delta t/\hbar}\mapsto \left(\bigotimes_{k=0}^{N_{\text{qubits}}-1}P_{k}(\theta_{g}/2)\right).
    \left[\bigotimes_{\substack{j=0,\\j(\mathrm{mod}\,2)=0}}^{L-1}\left(\bigotimes_{b=0}^{N-1}CP_{jN+b,(j+1)N+b}(-\theta_{g})\right).\right.\nonumber\\
    &\left.\left(\bigotimes_{\substack{b,c=0,\\ b<c}}^{N-1}(CP_{jN+b,jN+c}(\theta_{g})\otimes CP_{(j+1)N+b,(j+1)N+c}(\theta_{g})\otimes CP_{jN+b,(j+1)N+c}(-\theta_{g})\otimes CP_{(j+1)N+b,jN+c}(-\theta_{g}))\right)\right]\,,\label{eq:quartic-trotter-decompose}
\end{align}
where we denote, $\theta_{g}=\frac{g^{2}\delta t}{a\hbar}$, and $P_{k}(\theta)$ and $CP_{p,q}(\theta)$ are the single-qubit phase gate acting on $k$-th qubit and controlled-phase (CP) gate acting on $p$ and $q$-th qubits of the register, respectively.
The single Trotter step associated with the first-order Trotterization is given as,
\begin{equation}
    \hat{U}^{(1)}_{\text{Trotter}}(\delta t) = \hat{U}_{\text{even}}(\theta_{h}). \hat{U}_{\text{odd}}(\theta_{h}). \hat{U}_{\text{int}}(\theta_{g})\,,\label{eq:single-Trotter-1}
\end{equation}
where $\hat{U}_{\text{even}}$, $\hat{U}_{\text{odd}}$ and $\hat{U}_{\text{int}}$ are given by,
\begin{align}
    \hat{U}_{\text{even}}(\theta_{h}) & = \bigotimes_{\substack{k=0,\\k(\mathrm{mod}2N)=0}}^{N_{\text{qubits}}-1}U_{\text{hop}}(\theta_{h})\,,\label{eq:U-even}\\
    \hat{U}_{\text{odd}}(\theta_{h}) & = \bigotimes_{\substack{k=0,\\k(\mathrm{mod}2N)=N}}^{N_{\text{qubits}}-1}U_{\text{hop}}(\theta_{h})\,,\label{eq:U-odd}\\
    \hat{U}_{\text{int}}(\theta_{g}) & = \bigotimes_{\substack{k=0,\\k(\mathrm{mod}2N)=0}}^{N_{\text{qubits}}-1}U_{\text{int}}(\theta_{g})\,,\label{eq:U-int}
\end{align}
respectively. Here, both the unitary operators $U_{\text{hop}}(\theta_{h})$ and $U_{\text{int}}(\theta_{g})$ act on $2N$ qubits of the qubit register. The quantum circuit associated with the first-order Trotterization acting on consecutive fermionic lattice sites, $(2n,\,2n+1,\,2n+2,\,2n+3)$ is given in Fig.~\ref{fig:single-Trotter-step}.
Now, our qubit mapping introduced in Eq.~(\ref{eq:qubit-mapping}) translates the quadratic term (hopping) of the lattice Gross-Neveu model between fermionic sites $(2n,b)$ and $(2n+1,b)$ with $b$-th flavor where $b = 0,1,...,N-1$, into $(\text{next})^{N-1}$-nearest-neighbor qubit interactions. As the efficient and scalable Trotter circuits were already developed for one-dimensional Hamiltonians with next-nearest-neighbor interactions in Ref.~\cite{chowdhury2024enhancing, chowdhury2026quantum}, specifically in scenarios where qubit connectivity of a real quantum device is limited to nearest-neighbor connections, we adopt a similar strategy for the multi-flavor lattice Gross-Neveu Hamiltonian with long-range qubit interactions. In this respect, we incorporate SWAP gates in the circuit blocks to design a scalable Trotterization circuit as described below.

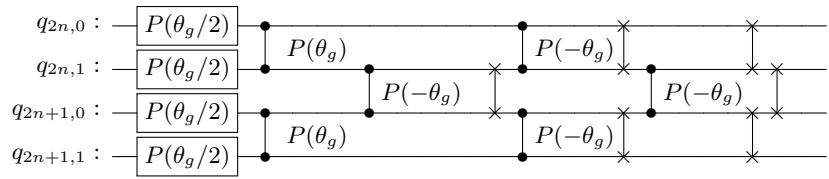
\begin{figure}[h!]
\centerline{
\Qcircuit @C=1.0em @R=0.2em @!R { \\
	 	\nghost{{q}_{2n,0} :  } & \lstick{{q}_{2n,0} :  } & \gate{P(\theta_{g}/2)} & \ctrl{1} & \dstick{\hspace{2.0em}P(\theta_{g})} \qw & \qw & \qw & \qw & \qw & \qw & \qw & \qw & \qw & \ctrl{1} & \dstick{\hspace{2.0em}P(-\theta_{g})} \qw & \qw & \qw & \qswap & \qw & \qw & \qw & \qw & \qswap & \qw & \qw & \qw\\
	 	\nghost{{q}_{2n,1} :  } & \lstick{{q}_{2n,1} :  } & \gate{P(\theta_{g}/2)} & \control \qw & \qw & \qw & \qw & \ctrl{1} & \dstick{\hspace{2.0em}P(-\theta_{g})} \qw & \qw & \qw & \qw & \qswap & \control \qw & \qw & \qw & \qw & \qswap \qwx[-1] & \ctrl{1} & \dstick{\hspace{2.0em}P(-\theta_{g})} \qw & \qw & \qw & \qswap \qwx[-1] & \qswap & \qw & \qw\\
	 	\nghost{{q}_{2n+1,0} :  } & \lstick{{q}_{2n+1,0} :  } & \gate{P(\theta_{g}/2)} & \ctrl{1} & \dstick{\hspace{2.0em}P(\theta_{g})} \qw & \qw & \qw & \control \qw & \qw & \qw & \qw & \qw & \qswap \qwx[-1] & \ctrl{1} & \dstick{\hspace{2.0em}P(-\theta_{g})} \qw & \qw & \qw & \qswap & \control \qw & \qw & \qw & \qw & \qswap & \qswap \qwx[-1] & \qw & \qw\\
	 	\nghost{{q}_{2n+1,1} :  } & \lstick{{q}_{2n+1,1} :  } & \gate{P(\theta_{g}/2)} & \control \qw & \qw & \qw & \qw & \qw & \qw & \qw & \qw & \qw & \qw & \control \qw & \qw & \qw & \qw & \qswap \qwx[-1] & \qw & \qw & \qw & \qw & \qswap \qwx[-1] & \qw & \qw & \qw\\
\\ }}
\caption{The quantum circuit of the corresponding unitary operator associated with the quartic interaction term acting on $(2n,2n+1)$ fermionic lattice sites for $N=2$ flavors.}
\label{fig:quartic-two-flavor}
\end{figure}

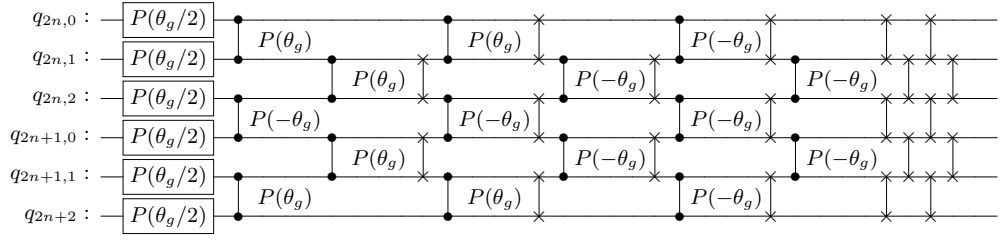
\begin{figure}[h!]
\centerline{
\scalebox{0.9}{
\Qcircuit @C=1.0em @R=0.2em @!R { \\
	 	\nghost{{q}_{2n,0} :  } & \lstick{{q}_{2n,0} :  } & \gate{P(\theta_{g}/2)} & \ctrl{1} & \dstick{\hspace{2.0em}P(\theta_{g})} \qw & \qw & \qw & \qw & \qw & \qw & \qw & \qw & \ctrl{1} & \dstick{\hspace{2.0em}P(\theta_{g})} \qw & \qw & \qw & \qswap & \qw & \qw & \qw & \qw & \qw & \ctrl{1} & \dstick{\hspace{2.0em}P(-\theta_{g})} \qw & \qw & \qw & \qswap & \qw & \qw & \qw & \qw & \qswap & \qw & \qswap & \qw & \qw & \qw\\
	 	\nghost{{q}_{2n,1} :  } & \lstick{{q}_{2n,1} :  } & \gate{P(\theta_{g}/2)} & \control \qw & \qw & \qw & \qw & \ctrl{1} & \dstick{\hspace{2.0em}P(\theta_{g})} \qw & \qw & \qw & \qswap & \control \qw & \qw & \qw & \qw & \qswap \qwx[-1] & \ctrl{1} & \dstick{\hspace{2.0em}P(-\theta_{g})} \qw & \qw & \qw & \qswap & \control \qw & \qw & \qw & \qw & \qswap \qwx[-1] & \ctrl{1} & \dstick{\hspace{2.0em}P(-\theta_{g})} \qw & \qw & \qw & \qswap \qwx[-1] & \qswap & \qswap \qwx[-1] & \qswap & \qw & \qw\\
	 	\nghost{{q}_{2n,2} :  } & \lstick{{q}_{2n,2} :  } & \gate{P(\theta_{g}/2)} & \ctrl{1} & \dstick{\hspace{2.0em}P(-\theta_{g})} \qw & \qw & \qw & \control \qw & \qw & \qw & \qw & \qswap \qwx[-1] & \ctrl{1} & \dstick{\hspace{2.0em}P(-\theta_{g})} \qw & \qw & \qw & \qswap & \control \qw & \qw & \qw & \qw & \qswap \qwx[-1] & \ctrl{1} & \dstick{\hspace{2.0em}P(-\theta_{g})} \qw & \qw & \qw & \qswap & \control \qw & \qw & \qw & \qw & \qswap & \qswap \qwx[-1] & \qswap & \qswap \qwx[-1] & \qw & \qw\\
	 	\nghost{{q}_{2n+1,0} :  } & \lstick{{q}_{2n+1,0} :  } & \gate{P(\theta_{g}/2)} & \control \qw & \qw & \qw & \qw & \ctrl{1} & \dstick{\hspace{2.0em}P(\theta_{g})} \qw & \qw & \qw & \qswap & \control \qw & \qw & \qw & \qw & \qswap \qwx[-1] & \ctrl{1} & \dstick{\hspace{2.0em}P(-\theta_{g})} \qw & \qw & \qw & \qswap & \control \qw & \qw & \qw & \qw & \qswap \qwx[-1] & \ctrl{1} & \dstick{\hspace{2.0em}P(-\theta_{g})} \qw & \qw & \qw & \qswap \qwx[-1] & \qswap & \qswap \qwx[-1] & \qswap & \qw & \qw\\
	 	\nghost{{q}_{2n+1,1} :  } & \lstick{{q}_{2n+1,1} :  } & \gate{P(\theta_{g}/2)} & \ctrl{1} & \dstick{\hspace{2.0em}P(\theta_{g})} \qw & \qw & \qw & \control \qw & \qw & \qw & \qw & \qswap \qwx[-1] & \ctrl{1} & \dstick{\hspace{2.0em}P(\theta_{g})} \qw & \qw & \qw & \qswap & \control \qw & \qw & \qw & \qw & \qswap \qwx[-1] & \ctrl{1} & \dstick{\hspace{2.0em}P(-\theta_{g})} \qw & \qw & \qw & \qswap & \control \qw & \qw & \qw & \qw & \qswap & \qswap \qwx[-1] & \qswap & \qswap \qwx[-1] & \qw & \qw\\
	 	\nghost{{q}_{2n+1,2} :  } & \lstick{{q}_{2n+2} :  } & \gate{P(\theta_{g}/2)} & \control \qw & \qw & \qw & \qw & \qw & \qw & \qw & \qw & \qw & \control \qw & \qw & \qw & \qw & \qswap \qwx[-1] & \qw & \qw & \qw & \qw & \qw & \control \qw & \qw & \qw & \qw & \qswap \qwx[-1] & \qw & \qw & \qw & \qw & \qswap \qwx[-1] & \qw & \qswap \qwx[-1] & \qw & \qw & \qw\\
\\ }}
}
\caption{The quantum circuit of the corresponding unitary operator associated with the quartic interaction term acting on $(2n,2n+1)$ fermionic lattice sites for $N=3$ flavors.}
\label{fig:quartic-three-flavor}
\end{figure}

\begin{figure}
    \centerline{\scalebox{0.9}{
\Qcircuit @C=1.0em @R=0.2em @!R { \\
	 	\nghost{{q}_{2n,0} :  } & \lstick{{q}_{2n,0} :  } & \gate{P(\theta_{g}/2)} & \ctrl{1} & \dstick{\hspace{2.0em}P(\theta_{g})} \qw & \qw & \qw & \qw & \qw & \qw & \qw & \qw & \ctrl{1} & \dstick{\hspace{2.0em}P(\theta_{g})} \qw & \qw & \qw & \qswap & \qw & \qw & \qw & \qw & \qw & \ctrl{1} & \dstick{\hspace{2.0em}P(-\theta_{g})} \qw & \qw & \qw & \qswap & \qw & \qw & \qw & \qw & \qw & \ctrl{1} & \dstick{\hspace{2.0em}P(\theta_{g})} \qw & \qw & \qw & \qswap & \qw & \qw & \qw & \qw & \qswap & \qw & \qswap & \qw & \qswap & \qw & \qw & \qw\\
	 	\nghost{{q}_{2n,1} :  } & \lstick{{q}_{2n,1} :  } & \gate{P(\theta_{g}/2)} & \control \qw & \qw & \qw & \qw & \ctrl{1} & \dstick{\hspace{2.0em}P(\theta_{g})} \qw & \qw & \qw & \qswap & \control \qw & \qw & \qw & \qw & \qswap \qwx[-1] & \ctrl{1} & \dstick{\hspace{2.0em}P(-\theta_{g})} \qw & \qw & \qw & \qswap & \control \qw & \qw & \qw & \qw & \qswap \qwx[-1] & \ctrl{1} & \dstick{\hspace{2.0em}P(-\theta_{g})} \qw & \qw & \qw & \qswap & \control \qw & \qw & \qw & \qw & \qswap \qwx[-1] & \ctrl{1} & \dstick{\hspace{2.0em}P(\theta_{g})} \qw & \qw & \qw & \qswap \qwx[-1] & \qswap & \qswap \qwx[-1] & \qswap & \qswap \qwx[-1] & \qswap & \qw & \qw\\
	 	\nghost{{q}_{2n,2} :  } & \lstick{{q}_{2n,2} :  } & \gate{P(\theta_{g}/2)} & \ctrl{1} & \dstick{\hspace{2.0em}P(\theta_{g})} \qw & \qw & \qw & \control \qw & \qw & \qw & \qw & \qswap \qwx[-1] & \ctrl{1} & \dstick{\hspace{2.0em}P(-\theta_{g})} \qw & \qw & \qw & \qswap & \control \qw & \qw & \qw & \qw & \qswap \qwx[-1] & \ctrl{1} & \dstick{\hspace{2.0em}P(-\theta_{g})} \qw & \qw & \qw & \qswap & \control \qw & \qw & \qw & \qw & \qswap \qwx[-1] & \ctrl{1} & \dstick{\hspace{2.0em}P(-\theta_{g})} \qw & \qw & \qw & \qswap & \control \qw & \qw & \qw & \qw & \qswap & \qswap \qwx[-1] & \qswap & \qswap \qwx[-1] & \qswap & \qswap \qwx[-1] & \qw & \qw\\
	 	\nghost{{q}_{2n,3} :  } & \lstick{{q}_{2n,3} :  } & \gate{P(\theta_{g}/2)} & \control \qw & \qw & \qw & \qw & \ctrl{1} & \dstick{\hspace{2.0em}P(-\theta_{g})} \qw & \qw & \qw & \qswap & \control \qw & \qw & \qw & \qw & \qswap \qwx[-1] & \ctrl{1} & \dstick{\hspace{2.0em}P(-\theta_{g})} \qw & \qw & \qw & \qswap & \control \qw & \qw & \qw & \qw & \qswap \qwx[-1] & \ctrl{1} & \dstick{\hspace{2.0em}P(-\theta_{g})} \qw & \qw & \qw & \qswap & \control \qw & \qw & \qw & \qw & \qswap \qwx[-1] & \ctrl{1} & \dstick{\hspace{2.0em}P(-\theta_{g})} \qw & \qw & \qw & \qswap \qwx[-1] & \qswap & \qswap \qwx[-1] & \qswap & \qswap \qwx[-1] & \qswap & \qw & \qw\\
	 	\nghost{{q}_{2n+1,0} :  } & \lstick{{q}_{2n+1,0} :  } & \gate{P(\theta_{g}/2)} & \ctrl{1} & \dstick{\hspace{2.0em}P(\theta_{g})} \qw & \qw & \qw & \control \qw & \qw & \qw & \qw & \qswap \qwx[-1] & \ctrl{1} & \dstick{\hspace{2.0em}P(-\theta_{g})} \qw & \qw & \qw & \qswap & \control \qw & \qw & \qw & \qw & \qswap \qwx[-1] & \ctrl{1} & \dstick{\hspace{2.0em}P(-\theta_{g})} \qw & \qw & \qw & \qswap & \control \qw & \qw & \qw & \qw & \qswap \qwx[-1] & \ctrl{1} & \dstick{\hspace{2.0em}P(-\theta_{g})} \qw & \qw & \qw & \qswap & \control \qw & \qw & \qw & \qw & \qswap & \qswap \qwx[-1] & \qswap & \qswap \qwx[-1] & \qswap & \qswap \qwx[-1] & \qw & \qw\\
	 	\nghost{{q}_{2n+1,1} :  } & \lstick{{q}_{2n+1,1} :  } & \gate{P(\theta_{g}/2)} & \control \qw & \qw & \qw & \qw & \ctrl{1} & \dstick{\hspace{2.0em}P(\theta_{g})} \qw & \qw & \qw & \qswap & \control \qw & \qw & \qw & \qw & \qswap \qwx[-1] & \ctrl{1} & \dstick{\hspace{2.0em}P(-\theta_{g})} \qw & \qw & \qw & \qswap & \control \qw & \qw & \qw & \qw & \qswap \qwx[-1] & \ctrl{1} & \dstick{\hspace{2.0em}P(-\theta_{g})} \qw & \qw & \qw & \qswap & \control \qw & \qw & \qw & \qw & \qswap \qwx[-1] & \ctrl{1} & \dstick{\hspace{2.0em}P(\theta_{g})} \qw & \qw & \qw & \qswap \qwx[-1] & \qswap & \qswap \qwx[-1] & \qswap & \qswap \qwx[-1] & \qswap & \qw & \qw\\
	 	\nghost{{q}_{2n+1,2} :  } & \lstick{{q}_{2n+1,2} :  } & \gate{P(\theta_{g}/2)} & \ctrl{1} & \dstick{\hspace{2.0em}P(\theta_{g})} \qw & \qw & \qw & \control \qw & \qw & \qw & \qw & \qswap \qwx[-1] & \ctrl{1} & \dstick{\hspace{2.0em}P(\theta_{g})} \qw & \qw & \qw & \qswap & \control \qw & \qw & \qw & \qw & \qswap \qwx[-1] & \ctrl{1} & \dstick{\hspace{2.0em}P(-\theta_{g})} \qw & \qw & \qw & \qswap & \control \qw & \qw & \qw & \qw & \qswap \qwx[-1] & \ctrl{1} & \dstick{\hspace{2.0em}P(\theta_{g})} \qw & \qw & \qw & \qswap & \control \qw & \qw & \qw & \qw & \qswap & \qswap \qwx[-1] & \qswap & \qswap \qwx[-1] & \qswap & \qswap \qwx[-1] & \qw & \qw\\
	 	\nghost{{q}_{2n+1,3} :  } & \lstick{{q}_{2n+1,3} :  } & \gate{P(\theta_{g}/2)} & \control \qw & \qw & \qw & \qw & \qw & \qw & \qw & \qw & \qw & \control \qw & \qw & \qw & \qw & \qswap \qwx[-1] & \qw & \qw & \qw & \qw & \qw & \control \qw & \qw & \qw & \qw & \qswap \qwx[-1] & \qw & \qw & \qw & \qw & \qw & \control \qw & \qw & \qw & \qw & \qswap \qwx[-1] & \qw & \qw & \qw & \qw & \qswap \qwx[-1] & \qw & \qswap \qwx[-1] & \qw & \qswap \qwx[-1] & \qw & \qw & \qw\\
\\ }}}
\caption{The quantum circuit of the corresponding unitary operator associated with the quartic interaction term acting on $(2n,2n+1)$ fermionic lattice sites for $N=4$ flavors.}
\label{fig:quartic-four-flavor}
\end{figure}
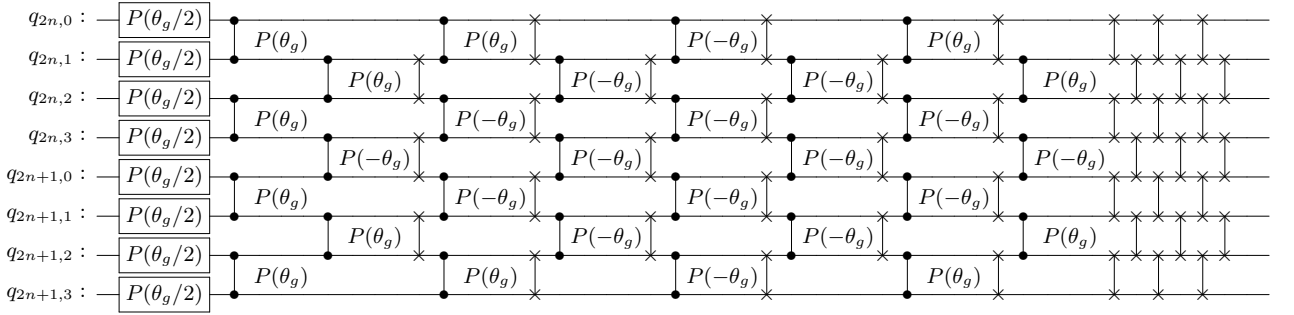

The quantum circuit associated with $U_{\text{hop}}(\theta_{h})$ acting on fermionic sites $(2n,\,2n+1)$ for $N=2$ flavors, is given as,
\[
\vcenter{
\Qcircuit @C=1.0em @R=1.0em @!R{
    \lstick{} & \multigate{3}{U_{\mathrm{hop}}(\theta_{h})} & \qw \\
    \lstick{} & \ghost{U_{\mathrm{hop}}(\theta_{h})}     & \qw \\
    \lstick{} & \ghost{U_{\mathrm{hop}}(\theta_{h})}     & \qw \\
    \lstick{} & \ghost{U_{\mathrm{hop}}(\theta_{h})}     & \qw \\
}
}
\hspace{1.0em}
=
\hspace{1.0em}
\vcenter{
\Qcircuit @C=1.0em @R=1.5em @!R {
    \nghost{q_{2n,0} : } & \lstick{q_{2n,0} : } & \qw & \multigate{1}{U_{\mathrm{XY-YX}}(\theta_{h})} & \qw & \qw \\
    \nghost{q_{2n,1} : } & \lstick{q_{2n,1} : } & \qswap & \ghost{U_{\mathrm{XY-YX}}(\theta_{h})} & \qswap & \qw \\
    \nghost{q_{2n+1,0} : } & \lstick{q_{2n+1,0} : } & \qswap \qwx[-1] & \multigate{1}{U_{\mathrm{XY-YX}}(\theta_{h})} & \qswap \qwx[-1] & \qw \\
    \nghost{q_{2n+1,1} : } & \lstick{q_{2n+1,1} : } & \qw & \ghost{U_{\mathrm{XY-YX}}(\theta_{h})} & \qw & \qw
}
}
\]
and for the same fermionic sites, in case of $N=3$ flavors, it is given as,
\[
\vcenter{
\Qcircuit @C=1.0em @R=1.0em @!R{
    \lstick{} & \multigate{5}{U_{\mathrm{hop}}(\theta_{h})} & \qw \\
    \lstick{} & \ghost{U_{\mathrm{hop}}(\theta_{h})}     & \qw \\
    \lstick{} & \ghost{U_{\mathrm{hop}}(\theta_{h})}     & \qw \\
    \lstick{} & \ghost{U_{\mathrm{hop}}(\theta_{h})}     & \qw \\
    \lstick{} & \ghost{U_{\mathrm{hop}}(\theta_{h})}     & \qw \\
    \lstick{} & \ghost{U_{\mathrm{hop}}(\theta_{h})}     & \qw \\
}
}
\hspace{1.0em}
=
\hspace{1.0em}
\vcenter{
\Qcircuit @C=1.0em @R=1.2em @!R {
    \nghost{q_{2n,0}:} & \lstick{q_{2n,0}:} & \qw & \qw & \multigate{1}{U_{\mathrm{XY-YX}}(\theta_{h})} & \qw & \qw & \qw & \qw \\
    \nghost{q_{2n,1}:} & \lstick{q_{2n,1}:} & \qw & \qswap & \ghost{U_{\mathrm{XY-YX}}(\theta_{h})} & \qswap & \qw & \qw & \qw \\
    \nghost{q_{2n,2}:} & \lstick{q_{2n,2}:} & \qswap & \qswap \qwx[-1] & \multigate{1}{U_{\mathrm{XY-YX}}(\theta_{h})} & \qswap \qwx[-1] & \qswap & \qw & \qw \\
    \nghost{q_{2n+1,0}:} & \lstick{q_{2n+1,0}:} & \qswap \qwx[-1] & \qswap & \ghost{U_{\mathrm{XY-YX}}(\theta_{h})} & \qswap & \qswap \qwx[-1] & \qw & \qw \\
    \nghost{q_{2n+1,1}:} & \lstick{q_{2n+1,1}:} & \qw & \qswap \qwx[-1] & \multigate{1}{U_{\mathrm{XY-YX}}(\theta_{h})} & \qswap \qwx[-1] & \qw & \qw & \qw \\
    \nghost{q_{2n+1,2}:} & \lstick{q_{2n+1,2}:} & \qw & \qw & \ghost{U_{\mathrm{XY-YX}}(\theta_{h})} & \qw & \qw & \qw & \qw
}
}
\]
where the SWAP-network facilitates the two-qubit unitary operator $U_{XY-YX}(\theta_{h})$ between the neighboring qubits. In fact, for $N$ flavors, the SWAP-network will have $2N$ SWAP gates and overall circuit-depth of $2(N-1)$ within a single circuit block $U_{\text{hop}}(\theta_{h})$.  In addition, the quantum circuit for the unitary operator $U_{XY-YX}(\theta_{h})=e^{-i\frac{\theta_{h}}{2}(XY-YX)}$ acting on two neighboring qubits can be expressed (for example, Ref.~\cite{Chai-Vincent-fermion}) as,
\[
\vcenter{
\Qcircuit @C=1.0em @R=1.0em @!R {
    \lstick{ } & \multigate{1}{U_{\mathrm{XY-YX}}(\theta_{h})} & \qw \\
    \lstick{} & \ghost{U_{\mathrm{XY-YX}}(\theta_{h})}     & \qw 
}
}
\hspace{1.0em}
=
\hspace{1.0em}
\vcenter{
\Qcircuit @C=1.0em @R=0.8em @!R { 
    \nghost{q_{k} : } & \lstick{q_{k} : } & \gate{R_X(-\frac{\pi}{2})} & \qw & \ctrl{1} & \gate{R_X(\theta_{h})}  & \ctrl{1} & \gate{R_X(\frac{\pi}{2})} & \qw & \qw \\
    \nghost{q_{k+1} : } & \lstick{q_{k+1} : } & \gate{R_Z(-\frac{\pi}{2})} & \gate{R_X(-\frac{\pi}{2})} & \targ & \gate{R_Z(\theta_{h})} & \targ & \gate{R_X(\frac{\pi}{2})} & \gate{R_Z(\frac{\pi}{2})} & \qw
}
}
\]
Similarly, the quartic interaction terms of the lattice Gross-Neveu Hamiltonian presented in Eq.~(\ref{eq:JW-quartic-0}) (or in Eq.~\ref{eq:JW-quartic}) acting on $(2n,2n+1)$ lattice sites across the same and distinct flavors lead to long-range qubit interactions. The circuit block representing the unitary operator $U_{\text{int}}(\theta_{g})$ acts on $2N$ qubits as shown in Fig.~\ref{fig:single-Trotter-step}, and requires $N(2N-1)$ two-qubit controlled-phase (CP) gates, resulting in a two-qubit circuit-depth of $2N$ within the block. Moreover, to facilitate these CP gates only acting on the neighboring qubits, we would require an overall $2(2N-1)(N-1)$ number of SWAP gates with a two-qubit circuit-depth of $4(N-1)$ within the circuit block of $U_{\text{int}}(\theta_{g})$. Hence, the total two-qubit circuit-depth associated with a single block $U_{\text{int}}(\theta_{g})$ turns out to be $2(3N-2)$. For example, the corresponding quantum circuit for $U_{\text{int}}(\theta_{g})$ acting on the fermionic sites $(2n,\,2n+1)$ with $N=2$ flavors, is given in Fig.~\ref{fig:quartic-two-flavor},
whereas, the quantum circuit of $U_{\text{int}}(\theta_{g})$ with $N=3$ flavors acting on the same fermionic sites is presented in Fig.~\ref{fig:quartic-three-flavor}.
On the other hand, for $N=4$ flavors the quantum circuit representing $U_{\text{int}}(\theta_{g})$ is shown in Fig.~\ref{fig:quartic-four-flavor}.
Note that, the quantum circuits for $U_{\text{int}}(\theta_{g})$ for $N$ flavors only act on the block of $2N$ qubits of the qubit register with total number of qubits $N_{\text{qubits}}=L N$, and do not depend on the total fermionic lattice sites $L$. Finally, the main advantage of the quantum circuit implementing the first-order Trotterization presented in this work is
that the total circuit-depth remains constant with increasing fermion sites $L$, and only depends on the number of flavors $N$, and Trotter steps $r$ which can be limited by the coherence time of quantum hardware.

The Trotter error associated with the first-order Trotterization for a fixed total simulation time $t$ scales as $O(t^{2}/r)$ with the number of $r$ Trotter steps, whereas for the second-order Trotterization, it scales as $O(t^{3}/r^{2})$ with $r$~\cite{Layden-Trotter-error}. The time evolution operator approximated by the second-order Trotterization is
\begin{equation}
    U^{(2)}(t)=\prod_{k=1}^{r}\hat{U}^{(2)}_{\text{Trotter}_{k}}(\delta t)\,,\label{eq:second-order-Trotter}
\end{equation}
where the single Trotter-step $\hat{U}^{(2)}_{\text{Trotter}}(\delta t)$ is given by
\begin{equation}
    \hat{U}^{(2)}_{\text{Trotter}}(\delta t)=\left(\prod_{s=1}^{M}e^{-iH_{s}\delta t/2\hbar}\right).\left(\prod_{l=1}^{M}e^{-i H_{M-l+1}\delta t/2\hbar}\right)\,.\label{eq:2nd-order-Trotter}
\end{equation}
By rearranging the Hamiltonian terms, we can rewrite Eq.~(\ref{eq:2nd-order-Trotter}) as,
\begin{align}
    \hat{U}^{(2)}_{\text{Trotter}}(\delta t)& =\hat{U}_{\text{even}}\left(\frac{\theta_{h}}{2}\right). \hat{U}_{\text{odd}}\left(\frac{\theta_{h}}{2}\right). \hat{U}_{\text{int}}\left(\frac{\theta_{g}}{2}\right).\hat{U}_{\text{int}}\left(\frac{\theta_{g}}{2}\right).\hat{U}_{\text{odd}}\left(\frac{\theta_{h}}{2}\right).\hat{U}_{\text{even}}\left(\frac{\theta_{h}}{2}\right)\nonumber\\
    & = \hat{U}_{\text{even}}\left(\frac{\theta_{h}}{2}\right).\hat{U}_{\text{odd}}\left(\frac{\theta_{h}}{2}\right). \hat{U}_{\text{int}}(\theta_{g}).\hat{U}_{\text{odd}}\left(\frac{\theta_{h}}{2}\right).\hat{U}_{\text{even}}\left(\frac{\theta_{h}}{2}\right)\,,\label{eq:second-Trotter-1}
\end{align}
where we combine two unitary operators $U_{\text{int}}(\theta_{g}/2)$ into $U_{\text{int}}(\theta_{g})$ in the single Trotter-step in second-order Trotterization. Furthermore, we can have
\begin{align}
    U^{(2)}(t) = &\left[\hat{U}_{\text{even}}\left(\frac{\theta_{h}}{2}\right).\hat{U}_{\text{odd}}\left(\frac{\theta_{h}}{2}\right). \hat{U}_{\text{int}}(\theta_{g}).\hat{U}_{\text{odd}}\left(\frac{\theta_{h}}{2}\right).\hat{U}_{\text{even}}\left(\frac{\theta_{h}}{2}\right)\right].\nonumber\\
    &\left[\hat{U}_{\text{even}}\left(\frac{\theta_{h}}{2}\right).\hat{U}_{\text{odd}}\left(\frac{\theta_{h}}{2}\right). \hat{U}_{\text{int}}(\theta_{g}).\hat{U}_{\text{odd}}\left(\frac{\theta_{h}}{2}\right).\hat{U}_{\text{even}}\left(\frac{\theta_{h}}{2}\right)\right]\,....\,\hat{U}_{\text{even}}\left(\frac{\theta_{h}}{2}\right)\,,\label{eq:second-Trotter-2}
\end{align}
where again combining $U_{\text{even}}(\theta_{h}/2)$ from two consecutive Trotter steps we have optimized second-order Trotterization,
\begin{align}
    \tilde{U}^{(2)}(t) = \hat{U}_{\text{even}}\left(\frac{\theta_{h}}{2}\right).\hat{U}_{\text{odd}}\left(\frac{\theta_{h}}{2}\right).\hat{U}_{\text{int}}(\theta_{g}).&\hat{U}_{\text{odd}}\left(\frac{\theta_{h}}{2}\right).\hat{U}_{\text{even}}(\theta_{h}).
    \hat{U}_{\text{odd}}\left(\frac{\theta_{h}}{2}\right). \hat{U}_{\text{int}}(\theta_{g}).\hat{U}_{\text{odd}}\left(\frac{\theta_{h}}{2}\right).\hat{U}_{\text{even}}(\theta_{h})\,....\nonumber\\
    &....\,\hat{U}_{\text{even}}\left(\theta_{h}\right).\hat{U}_{\text{odd}}\left(\frac{\theta_{h}}{2}\right). \hat{U}_{\text{int}}(\theta_{g}).\hat{U}_{\text{odd}}\left(\frac{\theta_{h}}{2}\right).\hat{U}_{\text{even}}\left(\frac{\theta_{h}}{2}\right)\,.\label{eq:second-Trotter-3}
\end{align}
Now for the first-order Trotterization, the total circuit depth $d^{(1)}$ with respect to the total number of Trotter steps $r$ scales as follows. For $N=2$ flavors, we have $d^{(1)}_{2}\sim 27\,r$ and for $N=3$ flavors, it is $d^{(1)}_{3}\sim 37\,r$. Now if we consider the second-order Trotterization given by Eq.~(\ref{eq:2nd-order-Trotter}), we have, for $N=2$ and $N=3$ flavors, the circuit depth as $d^{(2)}_{2}\sim 54\,r$ and $d^{(2)}\sim 74\,r$, respectively, which indicates that the circuit depth has doubled for a fixed number of Trotter steps as expected. On the other hand, for optimized second-order Trotterization presented in Eq.~(\ref{eq:second-Trotter-3}), the total circuit depths scale as $\tilde{d}^{(2)}_{2}\sim 37.73r^{0.98}$ and $\tilde{d}^{(2)}_{3}\sim 50.11 r^{0.98}$ for $N = 2$ and $N=3$ flavors, respectively, showcasing significant reduction in the total circuit depth compared to the second-order Trotterization.

\section{Localized Diagonal Operator Approximation}\label{sec:DOA}

In quantum computing, diagonal unitary operators are essential in variational quantum algorithms and quantum simulation. They define basis‑dependent phase rotations, which are central to Hamiltonian evolution and quantum arithmetic \cite{nielsen_chuang}.  
Hence, mapping long-range diagonal operators to a hardware topology with limited connectivity, like Nearest Neighbor (NN), is a critical task for implementing algorithms such as QAOA or Hamiltonian simulations (e.g., the Fermi-Hubbard model \cite{chowdhury2026quantum}).
Since diagonal operators commute with one another, we can rearrange them without affecting the final unitary, which allows for several elegant mapping and approximation strategies.
The most common method for implementing long-distance (beyond NN) diagonal interactions on a Linear Nearest Neighbor (LNN) architecture is the SWAP network.
This work also implements the SWAP network as described in Fig. \ref{fig:quartic-two-flavor}, \ref{fig:quartic-three-flavor}, and \ref{fig:quartic-four-flavor}.

In this work, we show that diagonal synthesis can be reformulated as a structured least‑squares projection in phase space.  
This connects to classical linear algebra via the Moore–Penrose pseudoinverse \cite{ben2003generalized}, and to hardware optimization through constrained Ansatz spaces \cite{cerezo2021variational}.
The key observation underlying this framework is that diagonal unitary synthesis reduces to a phase-matching problem in $\mathbb{R}^{2^n}$. Because diagonal two-qubit gates contribute phases additively and linearly in their parameters, the synthesis problem can be reformulated as a structured linear least-squares problem. This transforms unitary compilation into a linear algebraic projection problem in phase space.

\subsection{Motivation}

The quartic terms of the $2, 3,$ and $4$ flavors need long-distance (more than the next-nearest neighbor) interactions, and our initial circuit design for the long-distance interaction is inserting SWAP gates as shown in Fig. \ref{fig:quartic-two-flavor}, \ref{fig:quartic-three-flavor}, and \ref{fig:quartic-four-flavor}.
This design requires $4(N-1)$ SWAP-gate layers for the $N$- flavor system.
The quartic terms are composed of single-qubit phase gates, two-qubit phase gates (CP gates), and SWAP gates. 
Since the main challenge in implementing the quartic term is the two-qubit phase gate, we focus on its implementation.

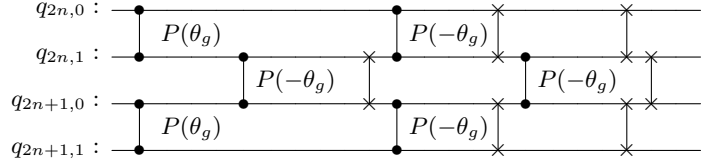
\begin{figure}[ht!]
\centering
\hspace{0.03\textwidth}
\[
\Qcircuit @C=1.0em @R=1.0em @!R {
	 	\nghost{{q}_{2n,0} :  } & \lstick{{q}_{2n,0} :  } &  \ctrl{1} & \dstick{\hspace{2.0em}P(\theta_{g})} \qw & \qw & \qw & \qw & \qw & \qw & \qw & \qw & \qw & \ctrl{1} & \dstick{\hspace{2.0em}P(-\theta_{g})} \qw & \qw & \qw & \qswap & \qw & \qw & \qw & \qw & \qswap & \qw & \qw & \qw\\
	 	\nghost{{q}_{2n,1} :  } & \lstick{{q}_{2n,1} :  } &  \control \qw & \qw & \qw & \qw & \ctrl{1} & \dstick{\hspace{2.0em}P(-\theta_{g})} \qw & \qw & \qw & \qw & \qswap & \control \qw & \qw & \qw & \qw & \qswap \qwx[-1] & \ctrl{1} & \dstick{\hspace{2.0em}P(-\theta_{g})} \qw & \qw & \qw & \qswap \qwx[-1] & \qswap & \qw & \qw\\
	 	\nghost{{q}_{2n+1,0} :  } & \lstick{{q}_{2n+1,0} :  } &  \ctrl{1} & \dstick{\hspace{2.0em}P(\theta_{g})} \qw & \qw & \qw & \control \qw & \qw & \qw & \qw & \qw & \qswap \qwx[-1] & \ctrl{1} & \dstick{\hspace{2.0em}P(-\theta_{g})} \qw & \qw & \qw & \qswap & \control \qw & \qw & \qw & \qw & \qswap & \qswap \qwx[-1] & \qw & \qw\\
	 	\nghost{{q}_{2n+1,1} :  } & \lstick{{q}_{2n+1,1} :  } &  \control \qw & \qw & \qw & \qw & \qw & \qw & \qw & \qw & \qw & \qw & \control \qw & \qw & \qw & \qw & \qswap \qwx[-1] & \qw & \qw & \qw & \qw & \qswap \qwx[-1] & \qw & \qw & \qw\\
}
\]
\caption{Target circuit for the diagonal operator approximation for the 2-flavor model. The circuit is constructed from CP gates for the two-flavor Gross--Neveu model. The CP and SWAP layer parts of the quartic term of the $2$ flavors. The one-qubit phase gates are removed from Fig. \ref{fig:quartic-two-flavor}.}
\label{fig:quartic-two-flavor-target}
\end{figure}

Figure \ref{fig:quartic-two-flavor-target} shows the two-qubit phase gate and the SWAP gates in Fig. \ref{fig:quartic-two-flavor}.
Since the two-qubit phase gates are diagonal operators and the SWAP gates change the pairs of interaction qubit, the operator described in Fig. \ref{fig:quartic-two-flavor-target} is diagonal too.
Our motivation is to approximate the target circuit (Fig. \ref{fig:quartic-two-flavor-target}) by the quantum circuit described in Fig. \ref{fig:quartic-two-flavor-appr} when one can not find a mathematical identity.
In other words, our method is finding the parameters, $x_0, x_1$, and $x_2$, minimizing the difference of the diagonal operators between the target circuit (Fig. \ref{fig:quartic-two-flavor-target}) and the Ansatz circuits (Fig. \ref{fig:quartic-two-flavor-appr}).
Since the CP gate and the RZZ gate are typical two-qubit diagonal gates, we study both cases.
Even though we suggest the Ansatz circuits described in Fig. \ref{fig:quartic-two-flavor-appr}, many variations are possible. For example, one can repeat the Ansatz circuits several times. Also, one can add single-qubit diagonal gates, such as Phase gates and Rz (z-axis rotation) gates.
When we tested these variations of the Ansatz, the approximation results were not better than the Ansatz in Fig. \ref{fig:quartic-two-flavor-appr}.

\begin{figure}[ht!]
\centering

\hspace{0.03\textwidth}
\begin{minipage}{0.3\textwidth}
\centering
\[
\Qcircuit @C=1.0em @R=1.8em @!R{
\lstick{q_0} & \ctrl{1} & \dstick{\hspace{2.0em}P(x_0)} \qw & \qw & \qw & \qw \\
\lstick{q_1} & \control \qw & \qw & \ctrl{1} & \dstick{\hspace{2.0em}P(x_2)} \qw & \qw \\
\lstick{q_2} & \ctrl{1} & \dstick{\hspace{2.0em}P(x_1)} \qw & \control \qw & \qw & \qw \\
\lstick{q_3} & \control \qw & \qw & \qw & \qw & \qw
}
\]
\caption*{(a) Ansatz circuit (CP gates).}
\end{minipage}
\begin{minipage}{0.3\textwidth}
\centering
\[
\Qcircuit @C=1.0em @R=1.8em @!R{
\lstick{q_0} & \ctrl{1} & \dstick{\hspace{2.0em}RZZ(x_0)} \qw & \qw & \qw & \qw \\
\lstick{q_1} & \control \qw & \qw & \ctrl{1} & \dstick{\hspace{2.0em}RZZ(x_2)} \qw & \qw \\
\lstick{q_2} & \ctrl{1} & \dstick{\hspace{2.0em}RZZ(x_1)} \qw & \control \qw & \qw & \qw \\
\lstick{q_3} & \control \qw & \qw & \qw & \qw & \qw
}
\]
\caption*{(b) Ansatz circuit (RZZ gates).}
\end{minipage}

\caption{Comparison of three 4-qubit circuits. (a) CP-based approximation Ansatz circuit obtained by applying the Diagonal Operator Approximation. (b) RZZ-based approximation ansatz circuit with equivalent interaction structure.}
\label{fig:quartic-two-flavor-appr}
\end{figure}

\subsection{Method}\label{sec:method}

We consider an $n$‑qubit diagonal unitary ($n = 2N$, $N$ flavors) of the target,
\begin{equation}\label{eq:target}
U_{\mathrm{target}} (\vec{\phi}(\theta_g))=
\mathrm{diag}(e^{i \phi_0},\ldots,e^{i \phi_{2^n-1}}),
\end{equation}
where $\vec{\phi}(\theta_g) = (\phi_0, \cdots, \phi_{2^n-1}) \in \mathbb{R}^{2^n}$. 
Since we have $1 = e^{i 0}$, the target operator can be described by the rotation angles.


Likewise, for the target operator, we construct the Ansatz diagonal operator from the Ansatz circuit using $m$ parameters of the Ansatz(e.g.,
CP or RZZ \cite{barenco1995elementary, muthukrishnan2000multivalued}. 
Note that we have $m = 2N - 1$ for $N$ flavors in the Ansatz design.
Let
\begin{equation}
\vec{x}=(x_0,\dots,x_{m-1}) \in \mathbb{R}^m
\end{equation}
denote the parameter vector corresponding to the Ansatz.
Now we define a vector function of the Ansatz as follows:

\begin{equation}
    \begin{aligned}
        \vec{\theta} : \mathbb{R}^m &\to \mathbb{R}^{2^n} \\
        \vec{x} &\mapsto \vec{\theta}(\vec{x})
    \end{aligned}
\end{equation}
where $\theta_k$ is $k$-th function value of $\vec{\theta}$ in $\mathbb{R}$, and we have $k=0, \cdots, 2^n - 1$.
Once one has the Ansatz circuit, defining the function $\vec{\theta}$ is straightforward because one can easily compute the phase of the Ansatz operator that is diagonal.
Hence, we have a diagonal operator of the Ansatz as follows:

\begin{equation}
\label{eq:unitary_ansatz}
U_{\mathrm{Ansatz}}(\vec{x})
=\mathrm{diag}(e^{i\theta_0(\vec{x})},\dots,e^{i\theta_{2^n-1}(\vec{x})}),
\end{equation}
where each \(\theta_k(\vec{x})\in\mathbb{R}\) and $\vec{\theta}(\vec{x}) = (\theta_0 (\vec{x}), \cdots, \theta_{2^n - 1} (\vec{x}))$.

Now, our goal is finding $\vec{x}$ satisfying
\begin{equation}\label{eq:global}
\min_{\vec{x} \in \mathbb{R}^m}
\left\| U_{\mathrm{target}}(\vec{\phi}(\theta_g)) - U_{\mathrm{Ansatz}}(\vec{\theta}(\vec{x})) \right\|
\end{equation}
for the given $\theta_g$ and the $L2$ norm $\left\|  \cdot \right\|$.
Instead of estimating $\left\| U_{\mathrm{target}}(\vec{\phi}(\theta_g)) - U_{\mathrm{Ansatz}}(\vec{\theta}(\vec{x})) \right\|$, we estimate $\left\| \vec{\phi}(\theta_g) - \vec{\theta}(\vec{x}) \right\|$. This means that we minimize the phase difference of two unitary operators, $U_{\mathrm{target}}(\vec{\phi}(\theta_g))$ and $U_{\mathrm{Ansatz}}(\vec{\theta}(\vec{x}))$.
This is justified when $\left\| \vec{\phi}(\theta_g) \right\|$ sufficiently small.
When we have sufficiently small $\left\| \vec{\phi}(\theta_g) \right\|$, we consider Taylor expansion around $0$.
Since we have $e^x \approx 1 + x$ around $0$ by the Taylor expansion, we have $\left\| U_{\mathrm{target}}(\vec{\phi}) - U_{\mathrm{Ansatz}}(\vec{\theta}(\vec{x})) \right\| \approx \left\| \vec{\phi}(\theta_g) - \vec{\theta}(\vec{x}) \right\|$ when we have $\vec{\phi}$ and $\vec{\theta}(\vec{x})$ which are close to $\vec{0}$.
Since $\phi$ is a linear vector function of $\theta_g$ presented in Fig. \ref{fig:quartic-two-flavor}, \ref{fig:quartic-three-flavor}, and \ref{fig:quartic-four-flavor}, and we have $\theta_g = \Delta t g^2 / a$ where $\Delta t$ is a Trotter step size.
Hence, we adopt sufficiently small $\Delta t$, and this induces small $\vec{\phi}$.

Note that $\vec{r} \in \mathbb{R}^{2^n}$ and $\vec{\theta}(\vec{x})$ is a linear combination of entries of $\vec{x}$ because the entries of $\vec{\theta}(\vec{x})$ are the phase of diagonal matrix, $U_{\mathrm{Ansatz}}(\vec{x})$.
Therefore, finding $\vec{x}$ minimizing $\left\| \vec{\phi}(\theta_g) - \vec{\theta}(\vec{x}) \right\|$ for given $\theta_g$ is a least squares problem.
Finally, we can rearrange the linear equations as follows:
\begin{equation}
\label{eq:linear_system_phase}
A \vec{x}= \vec{b},
\end{equation}
where \(A\in\mathbb{R}^{2^n\times m}\) and $\vec{b}$ includes entries of $\vec{\phi}$ and constants.
If we have $\left\| \vec{\phi}(\theta_g) - \vec{\theta}(\vec{x}) \right\| > 0$, the linear system,  $A \vec{x}= \vec{b}$, does not have a solution. In other words, when we have $\left\| \vec{\phi}(\theta_g) - \vec{\theta}(\vec{x}) \right\| > 0$, we have $\vec{b} \notin \mathrm{Col}(A)$, and we have to solve a least squares problem of $A \vec{x}= \vec{b}$.



\begin{figure}[!th]
    \centering
    \includegraphics[width=0.65\linewidth]{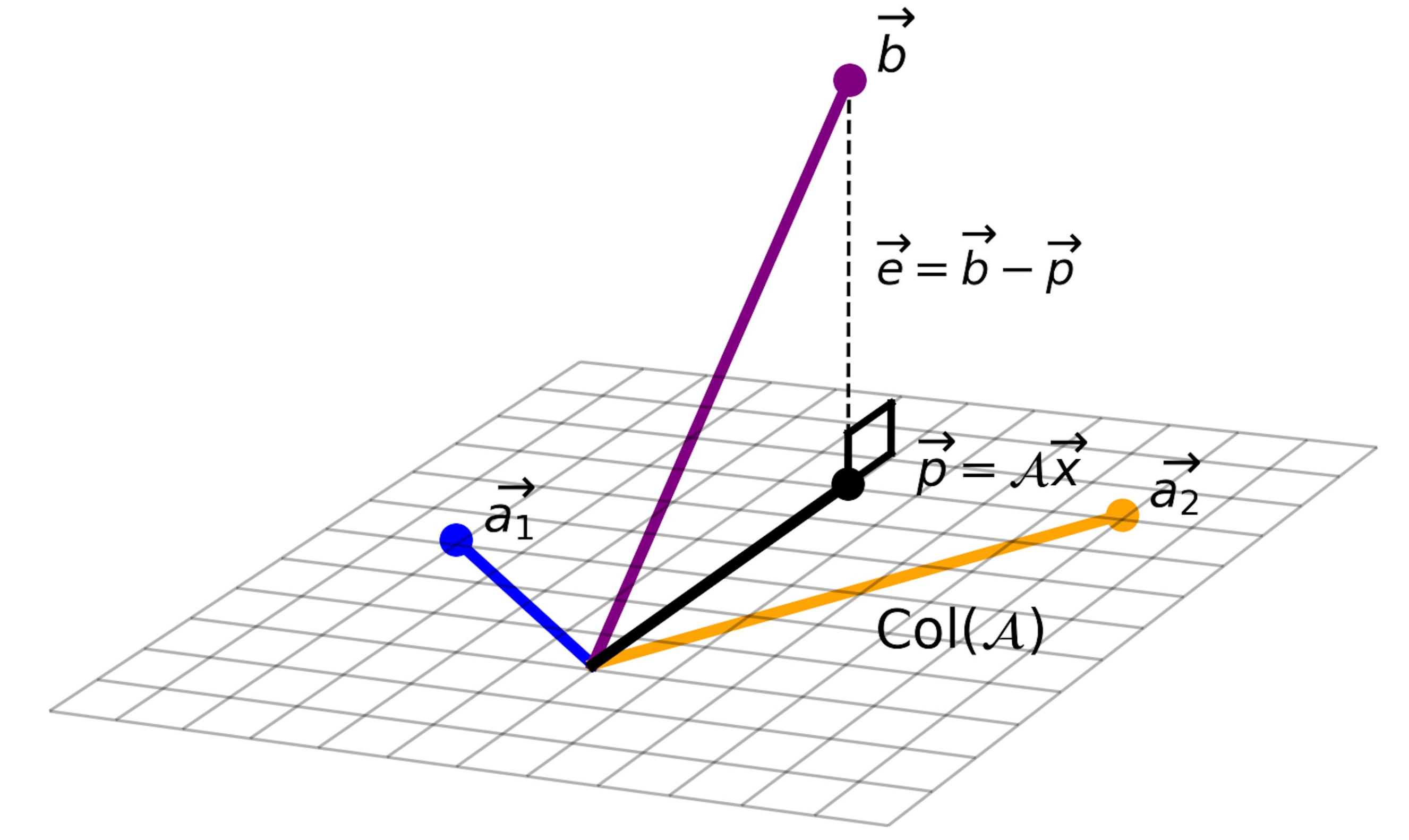}
\caption{
Geometric interpretation of a least squares problem.
The target vector $\vec{b}$ is projected onto the column space $\mathrm{Col}(A)$ spanned by $a_1$ and $a_2$ vectors,
yielding the closest realizable vector $\vec{p} = A\vec{x}$ for some $\vec{x}$, such that $\vec{e} = \vec{b} - \vec{p}$ is orthogonal to the subspace and
represents the approximation error due to limited Ansatz expressibility.
}    
\label{fig:gn_psuedo_inv}
\end{figure}

Figure~\ref{fig:gn_psuedo_inv} illustrates a geometric interpretation of a least squares problem.
The target phase vector $\vec{b} \in \mathbb{R}^{2^n}$ generally lies outside the column space
$\mathrm{Col}(A)$.
As a result, the linear system $A\vec{x} = \vec{b}$ is typically inconsistent, and no exact solution exists within the restricted parameterization.
The pseudoinverse resolves this by projecting $\vec{b}$ orthogonally onto $\mathrm{Col}(A)$,
yielding the closest realizable vector $\vec{p} = A\vec{x}$.
The residual vector $\vec{e} = \vec{b} - \vec{p}$ is orthogonal to the subspace, as indicated in Fig. \ref{fig:gn_psuedo_inv}, and quantifies the irreducible approximation error.
The solution $\vec{x}$ is given by $A^+ \vec{b}$ where the pseudoinverse $A^+$ is $(A^{T} A)^{-1} A^T$. Therefore, it corresponds to the minimum L2 norm parameter vector that best approximates the target phase within the Ansatz operator.






\subsection{Examples}
\subsubsection{2-flavor Model}



We apply the diagonal operator approximation to the two-flavor
Gross--Neveu evolution operator restricted to a fixed fermion-number sector.
In the form, the target unitary is diagonal in the computational basis as shown Fig. \ref{fig:quartic-two-flavor-target} and can be written as
\begin{equation}
\notag
U_{\mathrm{target}}(\vec{\phi}(\theta_g))
=
\mathrm{diag}
\left(
1,1,1,
e^{i\theta_g},
1,
e^{-i\theta_g},
e^{-i\theta_g},
e^{-i\theta_g},
1,
e^{-i\theta_g},
e^{-i\theta_g},
e^{-i\theta_g},
e^{i\theta_g},
e^{-i\theta_g},
e^{-i\theta_g},
e^{-2i\theta_g}
\right).
\end{equation}
We construct a diagonal Ansatz operator with three variational parameters $\vec{x} = (x_0, x_1, x_2)$ of the CP Ansatz as shown in Fig. \ref{fig:quartic-two-flavor-appr}a as follows:
\begin{equation}
\notag
U_{\mathrm{Ansatz}}(\vec{\theta}(x_0,x_1,x_2))
=
\mathrm{diag}
\left(
1,1,1,
e^{ix_0},
1,
1,
e^{ix_2},
e^{i(x_0 + x_2)},
1,
1,
1,
e^{ix_0},
e^{ix_1},
e^{ix_1},
e^{i (x_1 + x_2)},
e^{i (x_0 + x_1 + x_2)}
\right).
\end{equation}
Matching the diagonal phases reduces to a least squares problem, $\vec{\theta}(\vec{x}) = \vec{\phi}(\theta_g)$.
This form can be rearranged to the following linear system form after excluding the trivial identities $0=0$ and $1 = 1$:
\begin{equation}
\notag
\left\{
\begin{aligned}
 x_0 &= \theta_g \\
 0  &= -\theta_g \\
 x_2 &= -\theta_g \\
 x_0 + x_2 &= -\theta_g \\
 0 &= -\theta_g  \\
 0 &= -\theta_g  \\
 x_0 &= -\theta_g \\
 x_1 &= \theta_g \\
 x_1 &= -\theta_g \\
 x_1 + x_2 & = -\theta_g \\
 x_0 + x_1 + x_2 &= -2\theta
\end{aligned}
\right.
\end{equation}
These equations arise from the phase consistency conditions of the circuit for the $2$-flavor case. 
Collecting the above equations in matrix form yields $A \vec{x} = \vec{b}$ where $A \in \mathbb{R}^{11 \times 3}, 
\vec{x} = (x_0,x_1,x_2)^\top,
\vec{b} = (\phi_0,\phi_1,\dots,\phi_{2^n-1})^\top$.
Explicitly, the matrix, parameter vector, and target phase vector are
\begin{equation}
\notag
A =
\begin{pmatrix}
1 & 0 & 0 \\
0 & 0 & 0 \\
0 & 0 & 1 \\
1 & 0 & 1 \\
0 & 0 & 0 \\
0 & 0 & 0 \\
1 & 0 & 0 \\
0 & 1 & 0 \\
0 & 1 & 0 \\
0 & 1 & 1 \\
1 & 1 & 1
\end{pmatrix},
\qquad
x =
\begin{pmatrix}
x_0 \\ x_1 \\ x_2
\end{pmatrix},
\qquad
b =
\begin{pmatrix}
\theta \\
-\theta \\
-\theta \\
-\theta \\
-\theta \\
-\theta \\
-\theta \\
\theta \\
-\theta \\
-\theta \\
-2\theta
\end{pmatrix}.
\end{equation}

Since the system is overdetermined and not exactly solvable,
we determine the optimal parameters in the least-squares sense.
The Moore--Penrose pseudoinverse provides the minimum-norm solution
\begin{equation}
\notag
\vec{x}_{\mathrm{CP}}
=
A^{+} b,
\end{equation}
where the pseudoinverse is defined by $A^{+} = (A^{T}A)^{-1}A^{T}$, provided that $A^{T}A$ is invertible.
For the present circuit, the matrix $A$ has dimension
$11 \times 3$, corresponding to the number of computational
basis states contributing independent phase constraints.
The Moore--Penrose pseudoinverse can be evaluated analytically,
yielding
\begin{equation}
\notag
A^{+} =
\left(
\begin{array}{ccccccccccc}
\tfrac13 & 0 & -\tfrac16 & \tfrac16 & 0 & 0 & \tfrac13 & 0 & 0 & -\tfrac16 & \tfrac16 \\
0 & 0 & -\tfrac16 & -\tfrac16 & 0 & 0 & 0 & \tfrac13 & \tfrac13 & \tfrac16 & \tfrac16 \\
-\tfrac16 & 0 & \tfrac{5}{12} & \tfrac14 & 0 & 0 & -\tfrac16 & -\tfrac16 & -\tfrac16 & \tfrac14 & \tfrac1{12}
\end{array}
\right).
\end{equation}

\begin{figure}[t!]
    \centering
    \includegraphics[width=0.7\linewidth]{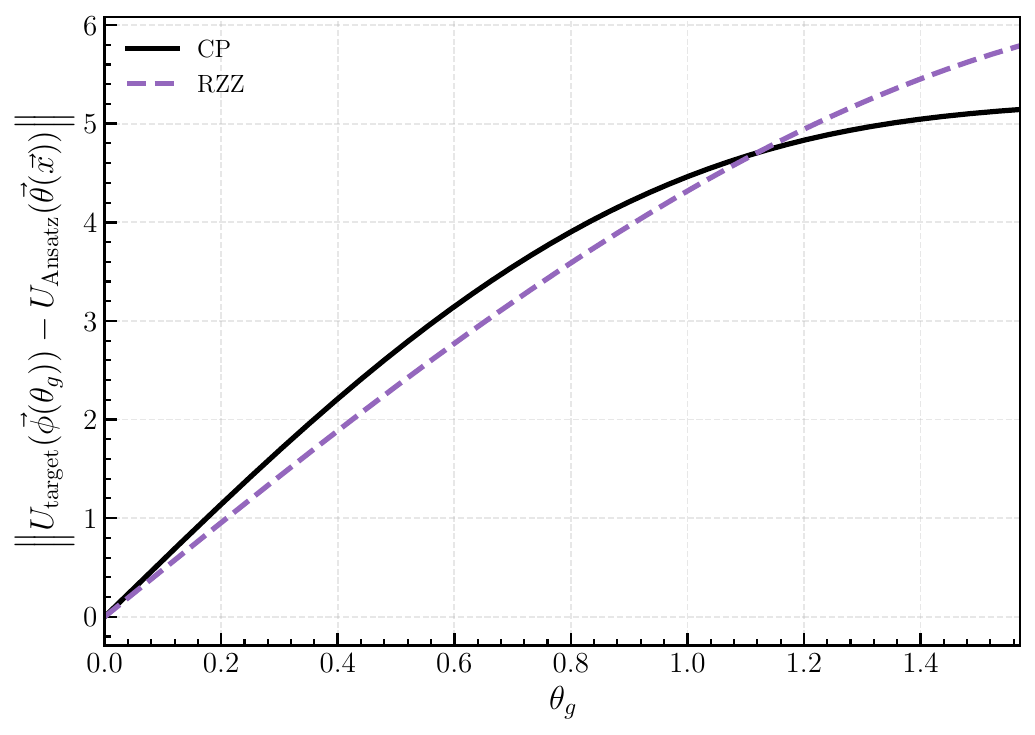}
    \caption{Comparison of the theoretical landscapes of the CP and RZZ Ansatze as a function of $\theta_g$. The black solid curve represents the CP Ansatz (Eq. (\ref{eq:Frob_norm_cp})), while the purple dashed curve corresponds to the RZZ Ansatz (Eq. (\ref{eq:Frob_norm_rzz})). Each curve shows the L2 norm between the target unitary $U_{\mathrm{target}}(\vec{\phi}(\theta_g))$ and the Ansatz unitary $U_{\mathrm{Ansatz}}(\vec{\theta}(\vec{x}))$.}
    \label{fig:l2_norm_2flavor}
\end{figure}

Evaluating the expression $\vec{x}_{\mathrm{CP}} = A^{+} b$ yields
\begin{equation}
\label{eq:x_CP_2f}
\vec{x}_{\mathrm{CP}} = 
\begin{pmatrix}
-\frac{\theta_g}{6} & -\frac{\theta_g}{6} & -\frac{13\theta_g}{12}
\end{pmatrix}^T.
\end{equation}
Thus, within the diagonal operator approximation, the optimal variational parameters are linear in $\theta_g$, with a nontrivial redistribution of phase weight among the three generators.
When we adopt the RZZ Ansatz of Fig. \ref{fig:quartic-two-flavor-appr}b, we have the following optimized parameters
\begin{equation}
\label{eq:x_Rzz_2f}
\vec{x}_{\mathrm{Rzz}} = 
\begin{pmatrix}
-\frac{\theta_g}{2} & -\frac{\theta_g}{2} & \frac{\theta_g}{2}
\end{pmatrix}^T.
\end{equation}

Finally, the L2 norms between the target and for each Ansatz ($\vec{x}_{\mathrm{CP}}$ and $\vec{x}_{\mathrm{Rzz}}$) operator have the following formulas:
\begin{flalign}
\label{eq:Frob_norm_cp}
\left\| U_{\mathrm{target}}(\vec{\phi}(\theta_g)) - U_{\mathrm{Ansatz}}(\vec{\theta}(\vec{x}_{CP})) \right\|^2 &= 
  22 -2 \cos \left( \frac{\theta_g}{12} \right) - 4 \cos \left( \frac{\theta_g}{4} \right) \\
\notag
& - 2 \cos \left( \frac{7\theta_g}{12} \right) - 4 \cos \left( \frac{5\theta_g}{6} \right) - 6 \cos (\theta_g) - 4 \cos \left( \frac{7\theta_g}{6} \right) 
\end{flalign}

\begin{flalign}
\label{eq:Frob_norm_rzz}
\left\| U_{\mathrm{target}}(\vec{\phi}(\theta_g)) - U_{\mathrm{Ansatz}}(\vec{\theta}(\vec{x}_{Rzz})) \right\|^2 &=32\!-\!4\cos(\frac{\theta_g}{4})\!-\!2\cos(\frac{3\theta_g}{8})\!-\!4\cos(\frac{5\theta_g}{8})\!-\!4\cos\theta_g\!-\!4\cos(\frac{5\theta_g}{4})\! 
\\  \notag
& -\!4\cos(\frac{11\theta_g}{8})\!-\!4\cos(\frac{15\theta_g}{8})\!-\!4\cos(\frac{9\theta_g}{4})\!-\!2\cos(\frac{29\theta_g}{8}) .
\end{flalign}

Note that both $\left\| U_{\mathrm{target}}(\vec{\phi}(\theta_g)) - U_{\mathrm{Ansatz}}(\vec{\theta}(\vec{x}_{Rzz})) \right\|$ and $\left\| U_{\mathrm{target}}(\vec{\phi}(\theta_g)) - U_{\mathrm{Ansatz}}(\vec{\theta}(\vec{x}_{Rzz})) \right\|$ converge to zero when $\theta_g$ approaches zero.
The $\left\| U_{\mathrm{target}}(\vec{\phi}(\theta_g)) - U_{\mathrm{Ansatz}}(\vec{\theta}(\vec{x})) \right\|$ with respect to $\theta_g$ ($[0, \pi/2]$) is plotted in Fig. \ref{fig:l2_norm_2flavor} which plots Eq. (\ref{eq:Frob_norm_cp}) and (\ref{eq:Frob_norm_cp}) with respect to $\theta_g$.
Figure \ref{fig:l2_norm_2flavor} also shows that the norms ($\left\| U_{\mathrm{target}}(\vec{\phi}(\theta_g)) - U_{\mathrm{Ansatz}}(\vec{\theta}(\vec{x}_{CP})) \right\|$ and $\left\| U_{\mathrm{target}}(\vec{\phi}(\theta_g)) - U_{\mathrm{Ansatz}}(\vec{\theta}(\vec{x}_{Rzz})) \right\|$) converge to zero as $\theta_g$ approaches zero.

\begin{figure}[ht!]
\centering

\hspace{0.03\textwidth}
\begin{minipage}{0.3\textwidth}
\centering
\[
\Qcircuit @C=1.5em @R=1.8em @!R{
\lstick{q_0} & \ctrl{1} & \dstick{\hspace{2.0em}P(-\frac{\theta_g}{6})} \qw & \qw & \qw & \qw \\
\lstick{q_1} & \control \qw & \qw & \ctrl{1} & \dstick{\hspace{2.0em}P(-\frac{13 \theta_g}{12})} \qw & \qw \\
\lstick{q_2} & \ctrl{1} & \dstick{\hspace{2.0em}P(-\frac{\theta_g}{6})} \qw & \control \qw & \qw & \qw \\
\lstick{q_3} & \control \qw & \qw & \qw & \qw & \qw
}
\]
\caption*{(a) CP-based approximation}
\end{minipage}
\begin{minipage}{0.3\textwidth}
\centering
\[
\Qcircuit @C=1.5em @R=1.8em @!R{
\lstick{q_0} & \ctrl{1} & \dstick{\hspace{2.0em}RZZ(-\frac{\theta_g}{2})} \qw & \qw & \qw & \qw \\
\lstick{q_1} & \control \qw & \qw & \ctrl{1} & \dstick{\hspace{2.0em}RZZ(\frac{\theta_g}{2})} \qw & \qw \\
\lstick{q_2} & \ctrl{1} & \dstick{\hspace{2.0em}RZZ(-\frac{\theta_g}{2})} \qw & \control \qw & \qw & \qw \\
\lstick{q_3} & \control \qw & \qw & \qw & \qw & \qw
}
\]
\caption*{(b) RZZ-based approximation}
\end{minipage}
\caption{The diagonal part approximation of the quartic part of the 2-flavor implementation.}
\label{fig:quartic-two-flavor-appr-run}
\end{figure}

Based on Eq. (\ref{eq:x_CP_2f}) and (\ref{eq:x_Rzz_2f}), we have the quantum circuit for the diagonal parts in the quartic term as described in Fig. \ref{fig:quartic-two-flavor-appr-run}.
The Ansatz parameters in Fig. \ref{fig:quartic-two-flavor-appr} are replaced by the values of Eq. (\ref{eq:x_CP_2f}) and (\ref{eq:x_Rzz_2f}), respectively.

\subsubsection{3-flavor Model}

We apply the diagonal operator approximation to the three-flavor Gross--Neveu evolution operator restricted to a fixed fermion-number sector.
In the form, the target unitary operator and the Ansatz unitary operators are visualized in Fig. \ref{fig:quartic-three-flavor-target} and \ref{fig:quartic-three-flavor-appr}, respectively.
\begin{figure}[h!]
\centering
\hspace{0.03\textwidth}
\[
\Qcircuit @C=1.0em @R=1.2em @!R { \\
	 	\nghost{{q}_{2n,0} :  } & \lstick{{q}_{2n,0} :  } &  \ctrl{1} & \dstick{\hspace{2.0em}P(\theta_{g})} \qw & \qw & \qw & \qw & \qw & \qw & \qw & \qw & \ctrl{1} & \dstick{\hspace{2.0em}P(\theta_{g})} \qw & \qw & \qw & \qswap & \qw & \qw & \qw & \qw & \qw & \ctrl{1} & \dstick{\hspace{2.0em}P(-\theta_{g})} \qw & \qw & \qw & \qswap & \qw & \qw & \qw & \qw & \qswap & \qw & \qswap & \qw & \qw & \qw\\
	 	\nghost{{q}_{2n,1} :  } & \lstick{{q}_{2n,1} :  } &  \control \qw & \qw & \qw & \qw & \ctrl{1} & \dstick{\hspace{2.0em}P(\theta_{g})} \qw & \qw & \qw & \qswap & \control \qw & \qw & \qw & \qw & \qswap \qwx[-1] & \ctrl{1} & \dstick{\hspace{2.0em}P(-\theta_{g})} \qw & \qw & \qw & \qswap & \control \qw & \qw & \qw & \qw & \qswap \qwx[-1] & \ctrl{1} & \dstick{\hspace{2.0em}P(-\theta_{g})} \qw & \qw & \qw & \qswap \qwx[-1] & \qswap & \qswap \qwx[-1] & \qswap & \qw & \qw\\
	 	\nghost{{q}_{2n,2} :  } & \lstick{{q}_{2n,2} :  } &  \ctrl{1} & \dstick{\hspace{2.0em}P(-\theta_{g})} \qw & \qw & \qw & \control \qw & \qw & \qw & \qw & \qswap \qwx[-1] & \ctrl{1} & \dstick{\hspace{2.0em}P(-\theta_{g})} \qw & \qw & \qw & \qswap & \control \qw & \qw & \qw & \qw & \qswap \qwx[-1] & \ctrl{1} & \dstick{\hspace{2.0em}P(-\theta_{g})} \qw & \qw & \qw & \qswap & \control \qw & \qw & \qw & \qw & \qswap & \qswap \qwx[-1] & \qswap & \qswap \qwx[-1] & \qw & \qw\\
	 	\nghost{{q}_{2n+1,0} :  } & \lstick{{q}_{2n+1,0} :  } &  \control \qw & \qw & \qw & \qw & \ctrl{1} & \dstick{\hspace{2.0em}P(\theta_{g})} \qw & \qw & \qw & \qswap & \control \qw & \qw & \qw & \qw & \qswap \qwx[-1] & \ctrl{1} & \dstick{\hspace{2.0em}P(-\theta_{g})} \qw & \qw & \qw & \qswap & \control \qw & \qw & \qw & \qw & \qswap \qwx[-1] & \ctrl{1} & \dstick{\hspace{2.0em}P(-\theta_{g})} \qw & \qw & \qw & \qswap \qwx[-1] & \qswap & \qswap \qwx[-1] & \qswap & \qw & \qw\\
	 	\nghost{{q}_{2n+1,1} :  } & \lstick{{q}_{2n+1,1} :  } &  \ctrl{1} & \dstick{\hspace{2.0em}P(\theta_{g})} \qw & \qw & \qw & \control \qw & \qw & \qw & \qw & \qswap \qwx[-1] & \ctrl{1} & \dstick{\hspace{2.0em}P(\theta_{g})} \qw & \qw & \qw & \qswap & \control \qw & \qw & \qw & \qw & \qswap \qwx[-1] & \ctrl{1} & \dstick{\hspace{2.0em}P(-\theta_{g})} \qw & \qw & \qw & \qswap & \control \qw & \qw & \qw & \qw & \qswap & \qswap \qwx[-1] & \qswap & \qswap \qwx[-1] & \qw & \qw\\
	 	\nghost{{q}_{2n+1,2} :  } & \lstick{{q}_{2n+2} :  } &  \control \qw & \qw & \qw & \qw & \qw & \qw & \qw & \qw & \qw & \control \qw & \qw & \qw & \qw & \qswap \qwx[-1] & \qw & \qw & \qw & \qw & \qw & \control \qw & \qw & \qw & \qw & \qswap \qwx[-1] & \qw & \qw & \qw & \qw & \qswap \qwx[-1] & \qw & \qswap \qwx[-1] & \qw & \qw & \qw\\
\\ }
\]
\caption{Target circuit for the diagonal operator approximation of the 3-flavor model. The circuit is constructed from CP gates for the two-flavor Gross--Neveu model. The CP and SWAP layer parts of the quartic term of the $3$ flavors. The one-qubit phase gates are removed from Fig. \ref{fig:quartic-three-flavor}.}
\label{fig:quartic-three-flavor-target}
\end{figure}
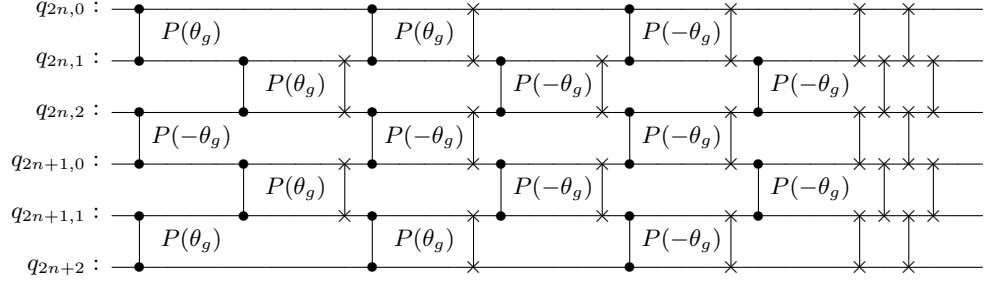

\begin{figure}[h!]
\centering
\begin{minipage}{0.43\textwidth}
\[
\Qcircuit @C=1.0em @R=1.8em @!R{
\lstick{q_0} & \ctrl{1} & \dstick{\hspace{2.0em}P(x_0)} \qw & \qw & \qw & \qw \\
\lstick{q_1} & \control \qw & \qw & \ctrl{1} & \dstick{\hspace{2.0em}P(x_3)} \qw & \qw \\
\lstick{q_2} & \ctrl{1} & \dstick{\hspace{2.0em}P(x_1)} \qw & \control \qw & \qw & \qw \\
\lstick{q_3} & \control \qw & \qw & \ctrl{1} & \dstick{\hspace{2.0em}P(x_4)} \qw & \qw \\
\lstick{q_4} & \ctrl{1} & \dstick{\hspace{2.0em}P(x_2)} \qw & \control \qw & \qw & \qw \\
\lstick{q_5} & \control \qw & \qw & \qw & \qw & \qw
}
\]
\caption*{(a) Ansatz circuit (CP gates)}
\end{minipage}
\begin{minipage}{0.43\textwidth}
\[
\Qcircuit @C=1.0em @R=1.8em @!R{
\lstick{q_0} & \ctrl{1} & \dstick{\hspace{2.0em}RZZ(x_0)} \qw & \qw & \qw & \qw \\
\lstick{q_1} & \control \qw & \qw & \ctrl{1} & \dstick{\hspace{2.0em}RZZ(x_3)} \qw & \qw \\
\lstick{q_2} & \ctrl{1} & \dstick{\hspace{2.0em}RZZ(x_1)} \qw & \control \qw & \qw & \qw \\
\lstick{q_3} & \control \qw & \qw & \ctrl{1} & \dstick{\hspace{2.0em}RZZ(x_4)} \qw & \qw \\
\lstick{q_4} & \ctrl{1} & \dstick{\hspace{2.0em}RZZ(x_2)} \qw & \control \qw & \qw & \qw \\
\lstick{q_5} & \control \qw & \qw & \qw & \qw & \qw
}
\]
\caption*{(b) Ansatz circuit (RZZ gates)}
\end{minipage}

\caption{Comparison of the two 6-qubit circuits. (a) CP-based Ansatz circuit obtained by applying the Diagonal Operator Approximation to approximate the target circuit. (b) RZZ-based Ansatz circuit obtained by applying the Diagonal Operator Approximation to approximate the target circuit.}
\label{fig:quartic-three-flavor-appr}
\end{figure}

In the same manner as in the two-flavor case, the formulae for the parameters of $\vec{x}_{\mathrm{CP}}$ and $\vec{x}_{\mathrm{Rzz}}$ are obtained as follows:
\begin{equation}
\label{eq:x_CP_3f}
\vec{x}_{\mathrm{CP}} = 
\begin{pmatrix}
-\frac{11\theta_g}{19} & -\frac{24\theta_g}{19} & -\frac{11\theta_g}{19} & \frac{\theta_g}{19}& \frac{\theta_g}{19}
\end{pmatrix}^T.
\end{equation}
\begin{equation}
\label{eq:x_Rzz_3f}
\vec{x}_{\mathrm{Rzz}} = 
\begin{pmatrix}
-\frac{\theta_g}{2} & \frac{\theta_g}{2} & -\frac{\theta_g}{2} & -\frac{\theta_g}{2} & -\frac{\theta_g}{2}
\end{pmatrix}^T.
\end{equation}

\subsubsection{4-flavor Model}

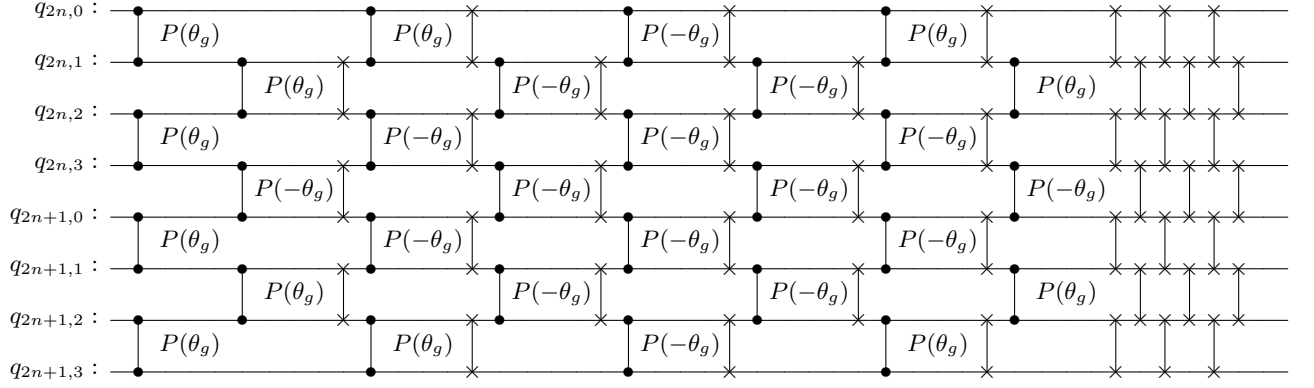
\begin{figure}[h!]
\centering
\hspace{0.03\textwidth}
\[
\Qcircuit @C=1.0em @R=1.2em @!R { \\
	 	\nghost{{q}_{2n,0} :  } & \lstick{{q}_{2n,0} :  } &  \ctrl{1} & \dstick{\hspace{2.0em}P(\theta_{g})} \qw & \qw & \qw & \qw & \qw & \qw & \qw & \qw & \ctrl{1} & \dstick{\hspace{2.0em}P(\theta_{g})} \qw & \qw & \qw & \qswap & \qw & \qw & \qw & \qw & \qw & \ctrl{1} & \dstick{\hspace{2.0em}P(-\theta_{g})} \qw & \qw & \qw & \qswap & \qw & \qw & \qw & \qw & \qw & \ctrl{1} & \dstick{\hspace{2.0em}P(\theta_{g})} \qw & \qw & \qw & \qswap & \qw & \qw & \qw & \qw & \qswap & \qw & \qswap & \qw & \qswap & \qw & \qw & \qw\\
	 	\nghost{{q}_{2n,1} :  } & \lstick{{q}_{2n,1} :  } &  \control \qw & \qw & \qw & \qw & \ctrl{1} & \dstick{\hspace{2.0em}P(\theta_{g})} \qw & \qw & \qw & \qswap & \control \qw & \qw & \qw & \qw & \qswap \qwx[-1] & \ctrl{1} & \dstick{\hspace{2.0em}P(-\theta_{g})} \qw & \qw & \qw & \qswap & \control \qw & \qw & \qw & \qw & \qswap \qwx[-1] & \ctrl{1} & \dstick{\hspace{2.0em}P(-\theta_{g})} \qw & \qw & \qw & \qswap & \control \qw & \qw & \qw & \qw & \qswap \qwx[-1] & \ctrl{1} & \dstick{\hspace{2.0em}P(\theta_{g})} \qw & \qw & \qw & \qswap \qwx[-1] & \qswap & \qswap \qwx[-1] & \qswap & \qswap \qwx[-1] & \qswap & \qw & \qw\\
	 	\nghost{{q}_{2n,2} :  } & \lstick{{q}_{2n,2} :  } &  \ctrl{1} & \dstick{\hspace{2.0em}P(\theta_{g})} \qw & \qw & \qw & \control \qw & \qw & \qw & \qw & \qswap \qwx[-1] & \ctrl{1} & \dstick{\hspace{2.0em}P(-\theta_{g})} \qw & \qw & \qw & \qswap & \control \qw & \qw & \qw & \qw & \qswap \qwx[-1] & \ctrl{1} & \dstick{\hspace{2.0em}P(-\theta_{g})} \qw & \qw & \qw & \qswap & \control \qw & \qw & \qw & \qw & \qswap \qwx[-1] & \ctrl{1} & \dstick{\hspace{2.0em}P(-\theta_{g})} \qw & \qw & \qw & \qswap & \control \qw & \qw & \qw & \qw & \qswap & \qswap \qwx[-1] & \qswap & \qswap \qwx[-1] & \qswap & \qswap \qwx[-1] & \qw & \qw\\
	 	\nghost{{q}_{2n,3} :  } & \lstick{{q}_{2n,3} :  } &  \control \qw & \qw & \qw & \qw & \ctrl{1} & \dstick{\hspace{2.0em}P(-\theta_{g})} \qw & \qw & \qw & \qswap & \control \qw & \qw & \qw & \qw & \qswap \qwx[-1] & \ctrl{1} & \dstick{\hspace{2.0em}P(-\theta_{g})} \qw & \qw & \qw & \qswap & \control \qw & \qw & \qw & \qw & \qswap \qwx[-1] & \ctrl{1} & \dstick{\hspace{2.0em}P(-\theta_{g})} \qw & \qw & \qw & \qswap & \control \qw & \qw & \qw & \qw & \qswap \qwx[-1] & \ctrl{1} & \dstick{\hspace{2.0em}P(-\theta_{g})} \qw & \qw & \qw & \qswap \qwx[-1] & \qswap & \qswap \qwx[-1] & \qswap & \qswap \qwx[-1] & \qswap & \qw & \qw\\
	 	\nghost{{q}_{2n+1,0} :  } & \lstick{{q}_{2n+1,0} :  } &  \ctrl{1} & \dstick{\hspace{2.0em}P(\theta_{g})} \qw & \qw & \qw & \control \qw & \qw & \qw & \qw & \qswap \qwx[-1] & \ctrl{1} & \dstick{\hspace{2.0em}P(-\theta_{g})} \qw & \qw & \qw & \qswap & \control \qw & \qw & \qw & \qw & \qswap \qwx[-1] & \ctrl{1} & \dstick{\hspace{2.0em}P(-\theta_{g})} \qw & \qw & \qw & \qswap & \control \qw & \qw & \qw & \qw & \qswap \qwx[-1] & \ctrl{1} & \dstick{\hspace{2.0em}P(-\theta_{g})} \qw & \qw & \qw & \qswap & \control \qw & \qw & \qw & \qw & \qswap & \qswap \qwx[-1] & \qswap & \qswap \qwx[-1] & \qswap & \qswap \qwx[-1] & \qw & \qw\\
	 	\nghost{{q}_{2n+1,1} :  } & \lstick{{q}_{2n+1,1} :  } &  \control \qw & \qw & \qw & \qw & \ctrl{1} & \dstick{\hspace{2.0em}P(\theta_{g})} \qw & \qw & \qw & \qswap & \control \qw & \qw & \qw & \qw & \qswap \qwx[-1] & \ctrl{1} & \dstick{\hspace{2.0em}P(-\theta_{g})} \qw & \qw & \qw & \qswap & \control \qw & \qw & \qw & \qw & \qswap \qwx[-1] & \ctrl{1} & \dstick{\hspace{2.0em}P(-\theta_{g})} \qw & \qw & \qw & \qswap & \control \qw & \qw & \qw & \qw & \qswap \qwx[-1] & \ctrl{1} & \dstick{\hspace{2.0em}P(\theta_{g})} \qw & \qw & \qw & \qswap \qwx[-1] & \qswap & \qswap \qwx[-1] & \qswap & \qswap \qwx[-1] & \qswap & \qw & \qw\\
	 	\nghost{{q}_{2n+1,2} :  } & \lstick{{q}_{2n+1,2} :  } &  \ctrl{1} & \dstick{\hspace{2.0em}P(\theta_{g})} \qw & \qw & \qw & \control \qw & \qw & \qw & \qw & \qswap \qwx[-1] & \ctrl{1} & \dstick{\hspace{2.0em}P(\theta_{g})} \qw & \qw & \qw & \qswap & \control \qw & \qw & \qw & \qw & \qswap \qwx[-1] & \ctrl{1} & \dstick{\hspace{2.0em}P(-\theta_{g})} \qw & \qw & \qw & \qswap & \control \qw & \qw & \qw & \qw & \qswap \qwx[-1] & \ctrl{1} & \dstick{\hspace{2.0em}P(\theta_{g})} \qw & \qw & \qw & \qswap & \control \qw & \qw & \qw & \qw & \qswap & \qswap \qwx[-1] & \qswap & \qswap \qwx[-1] & \qswap & \qswap \qwx[-1] & \qw & \qw\\
	 	\nghost{{q}_{2n+1,3} :  } & \lstick{{q}_{2n+1,3} :  } &  \control \qw & \qw & \qw & \qw & \qw & \qw & \qw & \qw & \qw & \control \qw & \qw & \qw & \qw & \qswap \qwx[-1] & \qw & \qw & \qw & \qw & \qw & \control \qw & \qw & \qw & \qw & \qswap \qwx[-1] & \qw & \qw & \qw & \qw & \qw & \control \qw & \qw & \qw & \qw & \qswap \qwx[-1] & \qw & \qw & \qw & \qw & \qswap \qwx[-1] & \qw & \qswap \qwx[-1] & \qw & \qswap \qwx[-1] & \qw & \qw & \qw\\
\\ }
\]
\caption{Target circuit for the diagonal operator approximation of the 4-flavor model. The circuit is constructed from CP gates for the two-flavor Gross--Neveu model. The CP and SWAP layer parts of the quartic term of the $4$ flavors. The one-qubit phase gates are removed from Fig. \ref{fig:quartic-four-flavor}.}
\label{fig:quartic-four-flavor-target}
\end{figure}

In this example, we apply the diagonal operator approximation to the four-flavor Gross--Neveu evolution operator restricted to a fixed fermion-number sector.
In the form, the target unitary operator and the Ansatz unitary operators are visualized in Fig. \ref{fig:quartic-four-flavor-target} and \ref{fig:quartic-four-flavor-appr}, respectively.

\begin{figure}[h!]
\centering
\hspace{0.1\textwidth}
\begin{minipage}{0.43\textwidth}
\[
\Qcircuit @C=1.0em @R=1.8em @!R{
\lstick{q_0} & \ctrl{1} & \dstick{\hspace{2.0em}P(x_0)} \qw & \qw & \qw & \qw \\
\lstick{q_1} & \control \qw & \qw & \ctrl{1} & \dstick{\hspace{2.0em}P(x_4)} \qw & \qw \\
\lstick{q_2} & \ctrl{1} & \dstick{\hspace{2.0em}P(x_1)} \qw & \control \qw & \qw & \qw \\
\lstick{q_3} & \control \qw & \qw & \ctrl{1} & \dstick{\hspace{2.0em}P(x_5)} \qw & \qw \\
\lstick{q_4} & \ctrl{1} & \dstick{\hspace{2.0em}P(x_2)} \qw & \control \qw & \qw & \qw \\
\lstick{q_5} & \control \qw & \qw & \ctrl{1} & \dstick{\hspace{2.0em}P(x_6)} \qw & \qw \\
\lstick{q_6} & \ctrl{1} & \dstick{\hspace{2.0em}P(x_3)} \qw & \control \qw & \qw & \qw \\
\lstick{q_7} & \control \qw & \qw & \qw & \qw & \qw
}
\]
\caption*{(a) Ansatz circuit (CP gates)}
\end{minipage}
\begin{minipage}{0.43\textwidth}
\[
\Qcircuit @C=1.5em @R=1.8em @!R{
\lstick{q_0} & \ctrl{1} & \dstick{\hspace{2.0em}RZZ(x_0)} \qw & \qw & \qw & \qw \\
\lstick{q_1} & \control \qw & \qw & \ctrl{1} & \dstick{\hspace{2.0em}RZZ(x_4)} \qw & \qw \\
\lstick{q_2} & \ctrl{1} & \dstick{\hspace{2.0em}RZZ(x_1)} \qw & \control \qw & \qw & \qw \\
\lstick{q_3} & \control \qw & \qw & \ctrl{1} & \dstick{\hspace{2.0em}RZZ(x_5)} \qw & \qw \\
\lstick{q_4} & \ctrl{1} & \dstick{\hspace{2.0em}RZZ(x_2)} \qw & \control \qw & \qw & \qw \\
\lstick{q_5} & \control \qw & \qw & \ctrl{1} & \dstick{\hspace{2.0em}RZZ(x_6)} \qw & \qw \\
\lstick{q_6} & \ctrl{1} & \dstick{\hspace{2.0em}RZZ(x_3)} \qw & \control \qw & \qw & \qw \\
\lstick{q_7} & \control \qw & \qw & \qw & \qw & \qw
}
\]
\caption*{(b) Ansatz circuit (RZZ gates)}
\end{minipage}
\caption{
Comparison of the two 8-qubit circuits. (a) CP-based Ansatz circuit obtained by applying the Diagonal Operator Approximation to approximate the target circuit. (b) RZZ-based Ansatz circuit obtained by applying the Diagonal Operator Approximation to approximate the target circuit.
}
\label{fig:quartic-four-flavor-appr}
\end{figure}

In the same manner as in the two-flavor and three-flavor cases, the formulae for the parameters of $\vec{x}_{\mathrm{CP}}$ and $\vec{x}_{\mathrm{Rzz}}$ are obtained as follows:
\begin{equation}
\label{eq:x_CP_4f}
\vec{x}_{\mathrm{CP}} = 
\begin{pmatrix}
-\frac{55\theta_g}{118} & -\frac{3\theta_g}{59} & -\frac{3\theta_g}{59} & -\frac{55\theta_g}{118}& -\frac{26\theta_g}{59} & -\frac{147\theta_g}{118} & -\frac{26\theta_g}{59}
\end{pmatrix}^T.
\end{equation}
\begin{equation}
\label{eq:x_Rzz_3f}
\vec{x}_{\mathrm{Rzz}} = 
\begin{pmatrix}
-\frac{\theta_g}{2} & -\frac{\theta_g}{2} & -\frac{\theta_g}{2} & -\frac{\theta_g}{2} & -\frac{\theta_g}{2} & \frac{\theta_g}{2} & -\frac{\theta_g}{2}
\end{pmatrix}^T.
\end{equation}

\subsection{Summary}

In summary, we have developed the Localized Diagonal Operator Approximation (LDOA) to circumvent the substantial gate overhead—specifically the $4(N-1)$ SWAP layers—required for multi-flavor quartic term interactions. By exploiting the diagonal nature of the target unitary, we demonstrated that complex long-distance interactions can be effectively compressed into a hardware-efficient Ansatz. This formulation is not restricted to the specific model considered here, but extends more broadly to diagonal quantum operators with long-range structure, independent of the underlying physical model, and is particularly well suited for implementation on quantum hardware with limited connectivity, such as superconducting architectures.
We proved the convergence of our LDOA method.
Also, our numerical results of two-, three-, and four-flavor confirm that optimizing the parameters of the Ansatz circuits yields a high-fidelity approximation. This approach significantly reduces the circuit depth while maintaining the essential diagonal dynamics, providing a scalable pathway for simulating large $N$-flavor systems on near-term quantum hardware.

\section{Trotterization circuits on IBM Quantum Computers}\label{sec:IBM-Trotterization}
To rigorously validate the accuracy and scalability of our diagonal operator approximation on physical superconducting quantum processors, it is imperative to benchmark the measurements from quantum devices against robust classical numerical methods. Because the dimensionality of the Hilbert space grows exponentially with the number of qubits, relying solely on exact classical diagonalization becomes unfeasible for larger systems. Therefore, our analysis is divided into two regimes: a moderately sized 20-qubit system where exact classical solutions are accessible, and a large-scale 108-qubit system that necessitates Tensor Network-based techniques for classical benchmarking. Throughout these experiments, we focus on the time evolution of the expectation value of the density-density correlator as defined in Eq.~(\ref{eq:density-density}) within the two-flavor Gross--Neveu model, comparing our proposed CP and RZZ approximation with classical benchmarks.
We implement the first-order Trotterization for the time evolution and the quantum error mitigation (QEM) methods detailed in Appendix~\ref {app:error_mitigations} to mitigate the influence of inherent quantum hardware imperfections, as well as hardware-efficient quantum circuit implementation methods \cite{zhang2024optimal, chowdhury2024enhancing}.

Within our quantum circuit implementation and subsequent experimental validation, the Localized Diagonal Operator Approximation (LDOA) plays a pivotal role in ensuring robust agreement with classical benchmarks. The primary challenge arises from the quartic terms associated with the 2, 3, and 4-flavor systems, which necessitate long-distance interactions extending beyond the next-nearest neighbor. In our baseline circuit design, these long-range dependencies were addressed through the systematic insertion of SWAP gates. However, this approach scales linearly with the system size, requiring $4(N-1)$ SWAP-gate layers for an $N$-flavor system, which significantly increases circuit depth and susceptibility to decoherence. Since these quartic terms are fundamentally composed of a complex interplay between single-qubit phase gates, two-qubit controlled phase gates, and SWAP operations, the LDOA is introduced as a strategic approximation. By streamlining these interactions, the LDOA effectively mitigates the computational overhead of the quartic terms, enabling more precise execution on near-term quantum hardware.
The diagonal operation parts in the quartic term (refer to Fig. \ref{fig:quartic-two-flavor}) are approximated by the circuits in  Fig. \ref{fig:quartic-two-flavor-appr-run} by the LDOA.

\begin{table}[h!]
    \centering
    \vspace{0.5cm}
    \begin{tabular}{c|c|c|c||c|c|c}
        \hline
        & \multicolumn{3}{c||}{~~~~~$L=10$ sites ($N = 20$ qubits)~~~~~} & \multicolumn{3}{c}{~~~~~$L=54$ sites ($N = 108$ qubits)~~~~~} \\
        \cline{2-7}
        Trotter Steps & Circuit depth & CZ depth & No. of CZ & Circuit depth & CZ depth & No. of CZ \\
        \hline
        \hline
        1 & 109 & 34 & 240 & 109 & 34 & 1340 \\
        \hline
        2 & 205 & 61 & 450 & 205 & 61 & 2518 \\
        \hline
        3 & 301 & 88 & 660 & 301 & 88 & 3696 \\
        \hline
        4 & 397 & 115 & 870 & 397 & 115 & 4874 \\
        \hline
        5 & 493 & 142 & 1080 & 493 & 142 & 6052 \\
        \hline
        6 & 589 & 169 & 1290 & 589 & 169 & 7230 \\
        \hline
        7 & 685 & 196 & 1500 & 685 & 196 & 8408 \\
        \hline
        8 & 781 & 223 & 1710 & 781 & 223 & 9586 \\
        \hline
    \end{tabular}
    \vspace{0.2cm}
    \caption{Circuit Statistics WITHOUT the LDOA (circuit in Fig. \ref{fig:quartic-two-flavor-target}). The circuit depth, CZ depth, and number of CZ gates with number of Trotter steps for the two flavor $L=10$ sites ($N=20$ qubits) and $L=54$ sites ($N=108$ qubits) on \texttt{ibm$\_$boston} after transpilation in Qiskit.}
    \label{tab:GN-statistics-1st-original}
\end{table}

\begin{table}[h!]
    \centering
    \vspace{0.5cm}
    \begin{tabular}{c|c|c|c||c|c|c}
        \hline
        & \multicolumn{3}{c||}{~~~~~$L=10$ sites ($N = 20$ qubits)~~~~~} & \multicolumn{3}{c}{~~~~~$L=54$ sites ($N = 108$ qubits)~~~~~} \\
        \cline{2-7}
        Trotter Steps & Circuit depth & CZ depth & No. of CZ & Circuit depth & CZ depth & No. of CZ \\
        \hline
        \hline
        1 & 62 & 18 & 120 & 62 & 18 & 692 \\
        \hline
        2 & 115 & 33 & 230 & 116 & 33 & 1330 \\
        \hline
        3 & 169 & 48 & 340 & 170 & 48 & 1968 \\
        \hline
        4 & 223 & 63 & 450 & 224 & 63 & 2606 \\
        \hline
        5 & 277 & 78 & 560 & 278 & 78 & 3244 \\
        \hline
        6 & 331 & 93 & 670 & 332 & 93 & 3882 \\
        \hline
        7 & 385 & 108 & 780 & 368 & 108 & 4520 \\
        \hline
        8 & 439 & 123 & 890 & 440 & 123 & 5158 \\
        \hline
    \end{tabular}
    \vspace{0.2cm}
    \caption{Circuit Statistics WITH the LDOA (RZZ Ansztz, circuit in Fig. \ref{fig:quartic-two-flavor-appr}b). The circuit depth, CZ depth, and number of CZ gates with number of Trotter steps for the two flavor $L=10$ sites ($N=20$ qubits) and $L=54$ sites ($N=108$ qubits) on \texttt{ibm$\_$boston} after transpilation in Qiskit.}
    \label{tab:GN-statistics-1st-DOA}
\end{table}

\begin{figure}[h!]
    \centering

    \subfloat[CZ Depth of 108-qubit system\label{fig:czdepth}]{
        \includegraphics[width=0.48\textwidth]{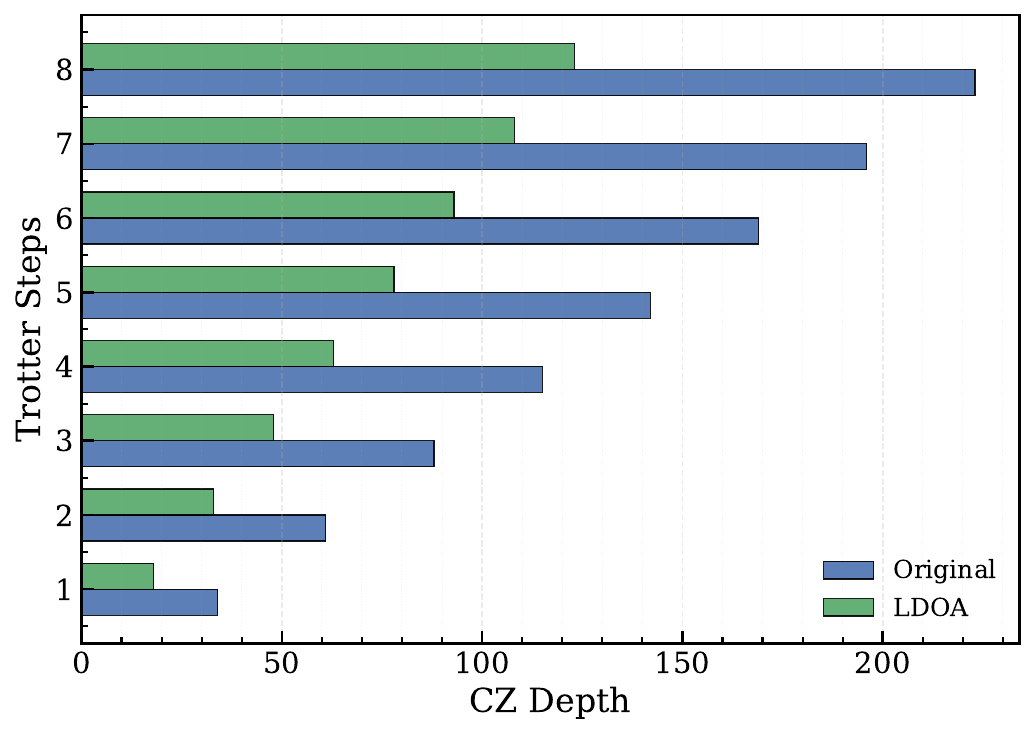}
    }
    \hfill
    \subfloat[Number of CZ gates of 108-qubit system\label{fig:cznumber}]{
        \includegraphics[width=0.48\textwidth]{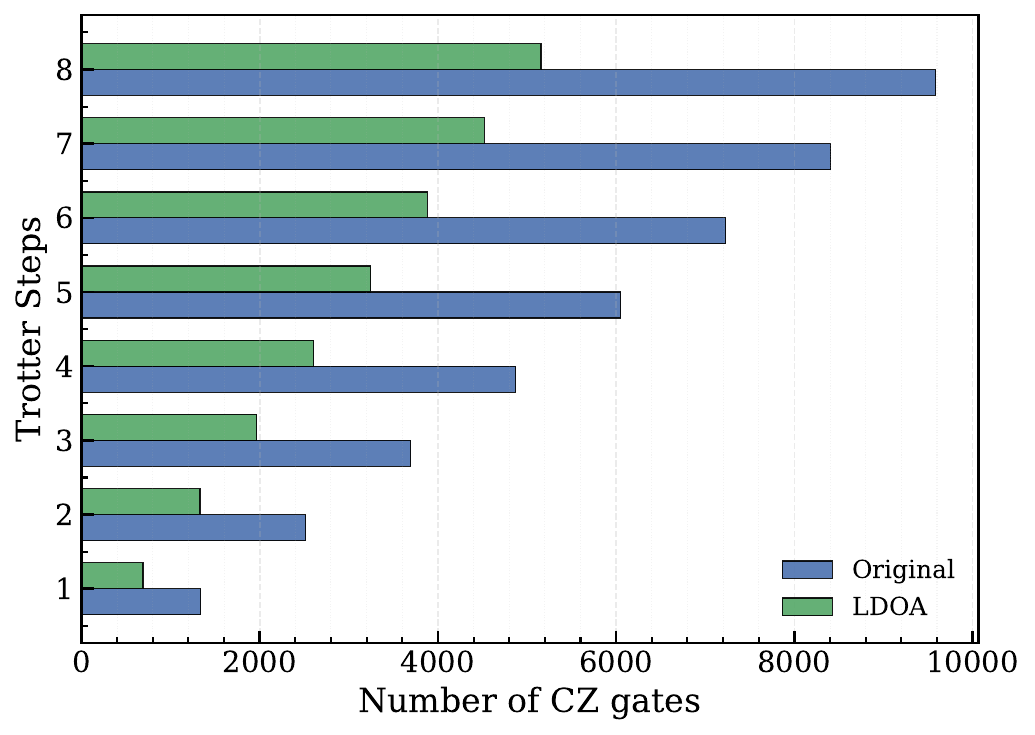}
    }

\caption{
Comparison of CZ gate depth and CZ gate count between the original circuit and the LDOA-optimized circuit.
Fig.~\ref{fig:czdepth} shows the CZ gate depth as a function of Trotter steps, while Fig.~\ref{fig:cznumber} shows the corresponding total number of CZ gates.
}    \label{fig:cz_doi}
\end{figure}

The comparative resource requirements for our quantum circuit implementations are detailed in Table~\ref{tab:GN-statistics-1st-original}, which covers the initial SWAP-based design, and Table~\ref{tab:GN-statistics-1st-DOA}, which outlines the Localized Diagonal Operator Approximation (LDOA) approach. These tables provide a comprehensive breakdown of total circuit depth, CZ depth, and the absolute CZ gate count. To further illustrate these metrics, Fig.~\ref{fig:cz_doi} provides a visual comparison of the CZ depth and gate count for the 108-qubit system (corresponding to a 2-flavor configuration with 54 sites). A cross-examination of the data in Table~\ref{tab:GN-statistics-1st-original} and Table~\ref{tab:GN-statistics-1st-DOA} reveals a strong alignment with the trends visualized in Fig.~\ref{fig:cz_doi}. In both instances, the LDOA implementation demonstrates a consistent and substantial reduction in both CZ depth and total gate count relative to the original SWAP-based construction. This trend remains robust across varying system sizes; for example, the 20-qubit system exhibits qualitatively identical behavior to the 108-qubit case. For the sake of conciseness and clarity in our presentation, we have focused our visual analysis in Fig.~\ref{fig:cz_doi} on the 108-qubit results. Furthermore, Fig.~\ref{fig:czdepth} highlights that the LDOA-enhanced circuit maintains a consistently lower CZ depth throughout all Trotter steps. This suggests that the advantages of LDOA are twofold: it not only diminishes the raw number of gates but also enables a more optimized spatial arrangement of two-qubit operations. This is corroborated by Fig.~\ref{fig:cznumber}, which shows a dramatic reduction in the total CZ gate count—specifically, the LDOA circuit requires approximately half the number of CZ gates compared to the original design across the entire simulation range. Collectively, these findings underscore the efficacy of the LDOA approach in compressing quantum resources and facilitating a hardware-efficient implementation that remains faithful to the underlying physical simulation.

\section{Experimental Results}\label{sec:Experimental-Results}
Here, we present the experimental results of the real-time dynamics in the lattice Gross-Neveu model from IBM quantum computers in the case of $N=2$. We benchmark the accuracy of our experimental results using a classical exact method for the small system size ($N_{\text{qubits}}=20$) and a classical approximation method based on Matrix product state (MPS) and time-dependent variational principle (TDVP) for the large-scale system ($N_{\text{qubits}}=108$). As our main goal is to validate the Trotterization circuit and its Localized Diagonal Operator Approximation (LDOA) presented in sections~\ref{sec:Trotterization}~and~\ref{sec:DOA}, we only consider single values of the lattice spacing as $a=1$ and the quartic coupling as $g = 0.5$ for simplicity.

\begin{figure}[t!]
\centering
\makebox[\textwidth][c]{%
\subfloat[CP Approxiamtion\label{fig:2flavor_cp}]{
\includegraphics[width=0.48\textwidth]{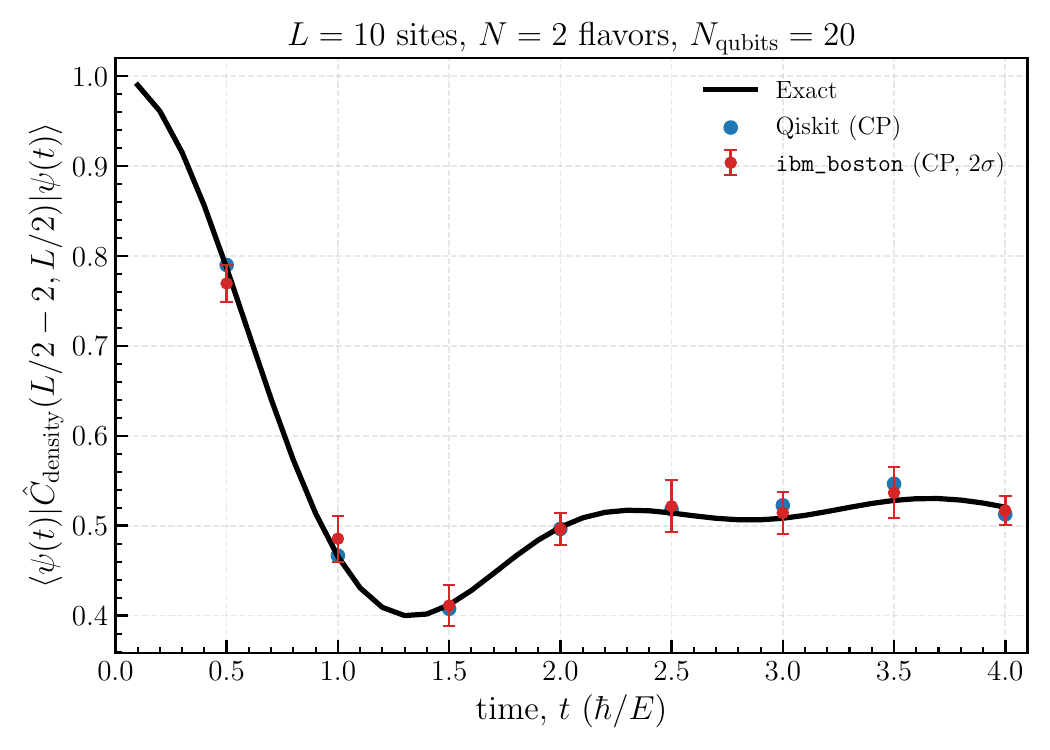}
}%
\subfloat[RZZ Approximation\label{fig:2flavor_rzz}]{
\includegraphics[width=0.48\textwidth]{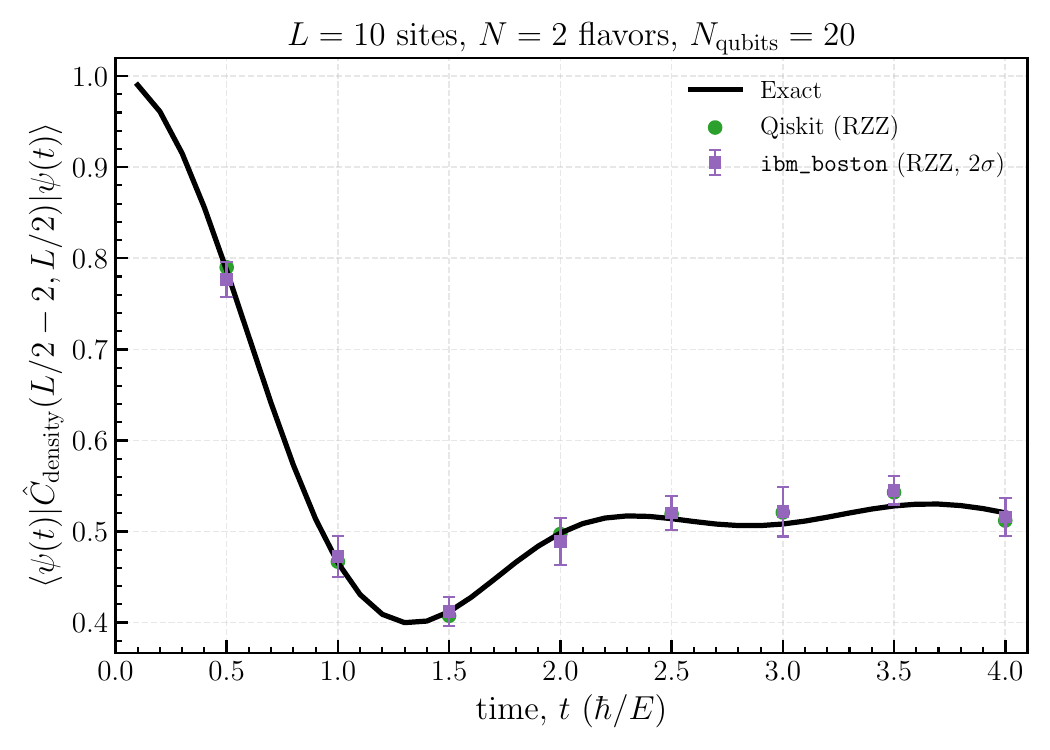}
}%
}
\caption{The time evolution of the density-density correlation in the two-flavor Gross-Neveu model with $ L=10$ sites with lattice spacing $a=1$ and quartic coupling $ g=0.5$. The left figure shows the results for the LDOA-implemented Trotterization circuit using the CP approximation with the QuSpin exact results (black solid lines), the noiseless Qiskit simulation (blue markers), and the experimental values from \texttt{ibm$\_$boston} (red markers) with $2\sigma$ uncertainty. The right figure shows the same using the RZZ approximation with the exact results (black solid line), noiseless Qiskit simulation (green markers), and experimental values (purple markers) from \texttt{ibm$\_$boston}.}
\label{fig:2flavor}
\end{figure}

\subsection{Two flavors with $L=10$ sites (20 qubits)}\label{sec:two-flavors-20-qubits}

We consider the expectation value of the density-density correlator, in other words the density-density correlation as outlined in Eq.~(\ref{eq:density-density}) for the time-evolved state $|\psi(t)\rangle$ for a system with two flavors and a lattice of size $L=10$, which corresponds to $N_{\text{qubits}}=20$. At this scale, the exact classical simulations remain feasible, enabling us to validate our Trotterization circuit design and its Localized Diagonal Operator Approximation (LDOA). Moreover, we can separate the Trotter error and the LDOA approximation errors from the hardware noise while working with a small system size.

We choose the initial state to be the product state $|\psi_{0}\rangle=|10011001100110011100\rangle$, which is not an eigenstate of the lattice Gross-Neveu Hamiltonian, so that it will have non-trivial time evolution. As depicted in Fig.~\ref{fig:2flavor}, the results show an excellent agreement between the exact calculations of the density-density correlation for representative sites $L/2$ and $L/2-2$, $\langle\psi(t)|\hat{C}_{\text{density}}(L/2-2,L/2)|\psi(t)\rangle$, performed using QuSpin~\cite{quspin}, and the noiseless simulation of LDOA-implemented Trotterized circuits in Qiskit~\cite{Qiskit}, for a simulation time up to $t=4$, measured in units of $\hbar/E$, where $E$ denotes the intrinsic energy scale of the lattice Gross-Neveu Hamiltonian. Additionally, the averaged density-density correlation obtained from five experimental runs at the \texttt{ibm$\_$boston} aligns well with both the exact results and the noiseless Qiskit simulations within $2\sigma$ of the statistical uncertainty.

\begin{figure}[t!]
    \centering
    \subfloat[CP Approxiamtion\label{fig:2flavor_cp}]{
    \includegraphics[width=0.48\textwidth]{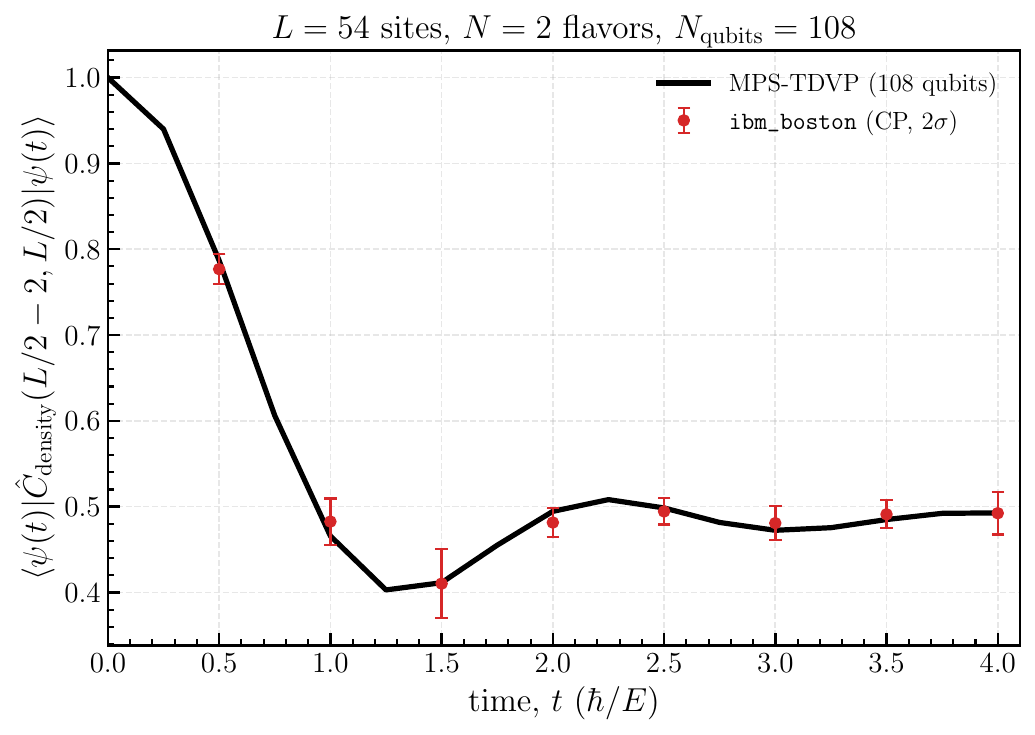}
    }
    \subfloat[RZZ Approximation\label{fig:2flavor_rzz}]{
    \includegraphics[width=0.48\textwidth]{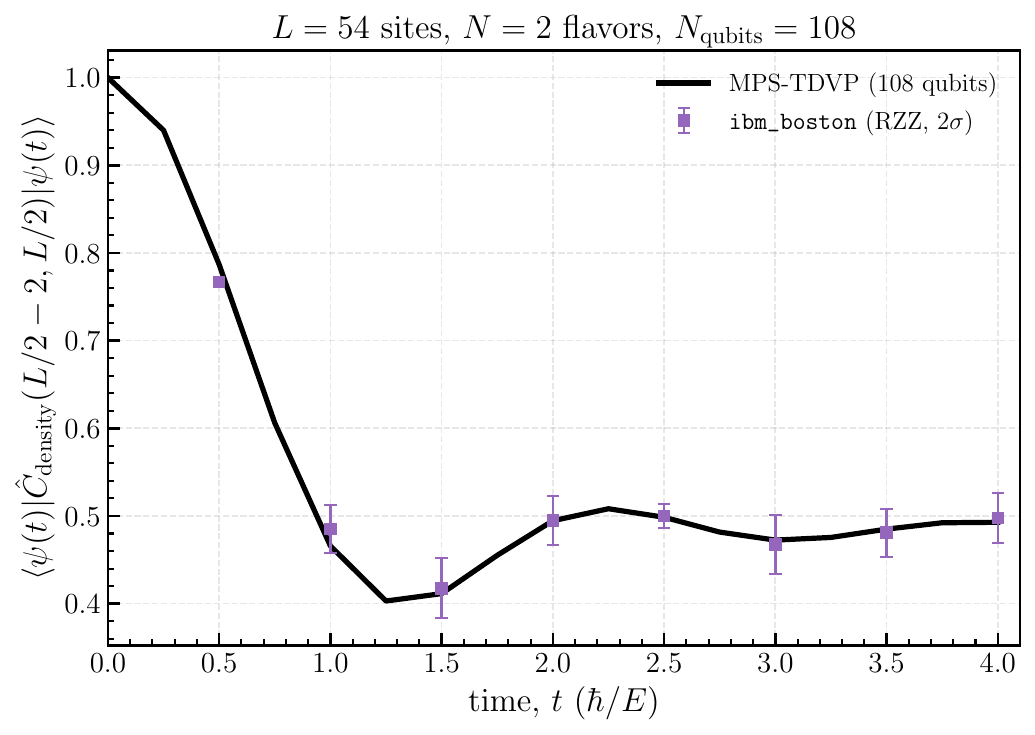}
    }
\caption{The time evolution of the density-density correlation in the two-flavor Gross-Neveu model with $ L=54$ sites ($N_{\text{qubits}}=108$ qubits) for $a=1$ and $ g=0.5$ with the MPS-TDVP results (black solid lines), and the experimental values (with $2\sigma$ uncertainties) from \texttt{ibm$\_$boston} for the LDOA-implemented Trotterization circuit using the CP approximation (red points in the left figure), and RZZ approximation (purple points in the right figure), respectively.
}    
\label{fig:large_2flavor}
\end{figure}

\subsection{Two flavors with $L=54$ sites (108 qubits)}\label{sec:two-flavors-108-qubits}

Now that we have established the reliability of our LDOA-implemented Trotterization circuits on a 20-qubit system, we extend our demonstration to a system of larger lattice sites $L=54$ with two flavors, involving $N_{\text{qubits}}= 108$. Simulating such system with more than hundred qubits using the exact method becomes intractable due to the exponential increase of the dimensionality of the corresponding Hilbert space of the system. Therefore, we have chosen the MPS-based time evolution method as our classical verifier for experimental results from IBM quantum computers involving 108 qubits. 

The Matrix Product State (MPS) is a mathematical structure consisting of a one-dimensional array of tensors, where each tensor represents a site or particle in a many-body system. The tensors are interconnected by bond indices, which can hold up to $\chi$ values as the bond dimension. The open indices of the tensors are linked to the physical degrees of freedom of the local Hilbert space associated with a particular fermionic site, taking up to $d$ values, which for the lattice Gross-Neveu model is $d = \text{dim}\mathcal{H}$ defined in Eq.~\ref{eq:dim-hj}. Although the MPS can encapsulate any quantum state of the many-body system, the bond dimension $\chi$ must increase exponentially with the system size to adequately represent all states in the Hilbert space. The time evolution of the MPS is determined using the time-dependent variational principle (TDVP)~\cite{TDVP-ref-1, TDVP-ref-2}, which is referred here as the MPS-TDVP approach, and is implemented using ITensor~\cite{itensor-1, itensor-2}. In the MPS-TDVP, we restrict the time evolution of the MPS to a particular tangent space, specified by bond dimensions, to the manifold of matrix product states. In other words, it projects the Hamiltonian's action on the MPS onto the MPS's tangent space, and then solves the time-dependent Schrödinger equation exclusively within that space. We simulate up to $t = 4$ with TDVP step-size $\delta_{\text{TDVP}}=0.25$ and $N_{\text{steps}}=16$ steps. Besides, we set the cutoff $\epsilon=10^{-8}$ and the maximum bond dimension $\chi_{\text{max}}=1200$ throughout the MPS-TDVP computation.

In Figure~\ref{fig:large_2flavor}, we present the results of LDOA-implemented Trotterization circuits involving CP approximation (Fig.~\ref{fig:large_2flavor}(a)) and RZZ approximation (Fig.~\ref{fig:large_2flavor}(b)) for the system size of 108 qubits. The initial state is again taken as $|\psi_{0}\rangle=|10011001... 1001... 10011100\rangle$, where `1001' is repeated 26 times from the left and has `1100' at the rightmost positions. We can see that the temporal variation of the averaged density-density correlation, $\langle\psi(t)|\hat{C}_{\text{density}}(L/2-2, L/2)|\psi(t)\rangle$ determined by \texttt{ibm$\_$boston} in both cases, agrees well with the corresponding values obtained by the MPS-TDVP computation. Finally, we would like to highlight that the LDOA implementation in our scalable Trotterization circuits enabled us to perform a large-scale quantum simulation of the lattice Gross-Neveu model, with a time limit set solely by the total coherence time of the IBM device.

\section{Entanglement Dynamics in the Gross-Neveu model}\label{sec:entropy}
Now that we have validated the Trotterization circuits for the quantum simulation of the Gross-Neveu model on IBM quantum computers across different system sizes, as presented in section~\ref{sec:Experimental-Results}. Therefore, by utilizing our Trotterization circuits, we can also probe the dynamics of entanglement entropy in the Gross-Neveu model under time evolution using IBM quantum computers. Entanglement entropy is a vital diagnostic tool in quantum systems, providing important insights into the system's fundamental quantum structure and its nonlocal correlations~\cite{horodecki}. If the Jordan-Wigner-transformed lattice Gross-Neveu model with total number of qubits given by $N_{\text{qubits}}$, is in a pure state $|\psi\rangle$ with density matrix $\rho = |\psi\rangle\langle\psi|$, we partition the system into two subsystems: $A$ (with number of qubits given by $L_{\text{qubits}}$) and $B$ (with $M_{\text{qubits}} = N_{\text{qubits}} - L_{\text{qubits}}$). Now, the R\'enyi entropy of order $n$ associated with subsystem $A$ is defined as:
\begin{equation}
S^{(n)}_{A} = \frac{1}{1-n} \log \left[ \mathrm{Tr}(\rho_{A}^{n}) \right],
\label{eq:n-Renyi entropy}
\end{equation}
where $\rho_{A} = \mathrm{Tr}_{B}(\rho)$ is the reduced density matrix obtained by tracing out subsystem $B$. In this study, we only focus on measuring the second-order R\'enyi entropy ($S^{(2)}_{A}$). Besides, when $n \to 1$, the R\'enyi entropy converges to the von Neumann entropy:
\begin{equation}
S^{vN}_{A} = -\mathrm{Tr} \left( \rho_{A} \log \rho_{A} \right).
\label{eq:von neumann entropy}
\end{equation}

Despite its theoretical significance, measuring entanglement entropy on current quantum hardware remains challenging. Although Quantum State Tomography (QST) provides a method for reconstructing the density matrix $\rho$ of a quantum system and thus the entanglement entropy, its applicability is limited to very small systems due to the exponential increase in the number of required measurements as the system size grows. To overcome this limitation, we adopt a more pragmatic method for the near-term quantum hardware, namely the randomized measurement protocol.

\subsection{Randomized Measurement Protocol}\label{sec:randomized-protocol}

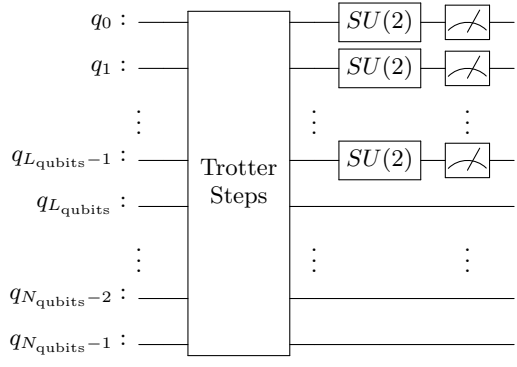
\begin{figure}[t!]
\[
\Qcircuit @C=1.0em @R=0.3em @!R{
\lstick{q_{0}: }   & \qw & \multigate{7}{ \makecell{\mathrm{Trotter}\\\mathrm{Steps}} } & \qw & \gate{SU(2)} & \meter &\qw\\
\lstick{q_{1}: }   & \qw & \ghost{ \makecell{\mathrm{Trotter}\\\mathrm{Steps}} }        & \qw & \gate{SU(2)} & \meter & \qw\\
\vdots & & & \vdots & &  \vdots\\
\lstick{q_{L_{\text{qubits}}-1}: } & \qw & \ghost{ \makecell{\mathrm{Trotter}\\\mathrm{Steps}} }        & \qw & \gate{SU(2)} & \meter & \qw \\
\lstick{q_{L_{\text{qubits}}}: } & \qw & \ghost{ \makecell{\mathrm{Trotter}\\\mathrm{Steps}} }        & \qw & \qw & \qw & \qw \\
\vdots & & & \vdots & &  \vdots\\
\lstick{q_{N_{\text{qubits}}-2}: } & \qw & \ghost{ \makecell{\mathrm{Trotter}\\\mathrm{Steps}} }        & \qw & \qw & \qw & \qw \\
\lstick{q_{N_{\text{qubits}}-1}: } & \qw & \ghost{ \makecell{\mathrm{Trotter}\\\mathrm{Steps}} }        & \qw & \qw & \qw    & \qw 
}
\]
\caption{The quantum circuit associated with the randomized measurement
(RM) protocol. Here, the local Haar-random unitaries are applied on the 
subsystem A whose Rényi entropy is being measured. 
}
\label{fig:randomized-protocol}
\end{figure}

The Randomized Measurement (RM) protocol \cite{Brydges-randomized, Enk-Beenakker, Elben-1, Vermersch-1, Elben-2, Rath, Elben-3} estimates R\'enyi entropy through statistical correlations of measurement outcomes in randomly sampled bases. 
As illustrated in the schematic circuit in Fig. \ref{fig:randomized-protocol}, the process involves several key stages. First, the highly entangled state $\rho$ is prepared via Trotterized time evolution of the initial product state $|\psi_{0}\rangle$. Then we apply a product of single-qubit random unitaries on subsystem A:
\begin{equation}
\hat{U}_{a} = U^{(2)}_{1} \otimes U^{(2)}_{2} \otimes \dots \otimes U^{(2)}_{L_{\text{qubits}}}\,,
\label{eq:single-qubit-local-unitary}
\end{equation}
where each $U^{(2)}_{i}$ is independently sampled from the Circular Unitary Ensemble (CUE) of $SU(2)$. Following the application of $\hat{U}_{a}$, we measure the qubits in the computational basis ($Z$-basis). For each $\hat{U}_{a}$, one conducts repeated measurements to obtain the statistics for estimating the occupation probabilities, $P_{\hat{U}_{a}}(\mathbf{s}_{A})=\mathrm{Tr}[\hat{U}_{a}\rho\hat{U}^{\dagger}_{a}|\mathbf{s}_{A}\rangle\langle \mathbf{s}_{A}|]$ of computational basis states $|\mathbf{s}_{A}\rangle = |s_{1},s_{2},...,s_{L_{\text{qubits}}}\rangle$ with $s_{i}=0,1$. Afterward, the entire process is repeated for $N_{U}$ different randomly drawn instances of $\hat{U}_{a}$.
After determining the set of outcome probabilities $P_{\hat{U}_{a}}(\mathbf{s}_{A})$ of the computational basis states $|\mathbf{s}_{A}\rangle$ for one instance of random unitaries, $\hat{U}_{a}$, one computes the following quantity,
\begin{equation}
    X_{a} = 2^{L_{\text{qubits}}}\sum_{\mathbf{s}_{A},\mathbf{s}'_{A}}(-2)^{-D[\mathbf{s}_{A},\mathbf{s}'_{A}]}P_{\hat{U}_{a}}(\mathbf{s}_{A})P_{\hat{U}_{a}}(\mathbf{s}'_{A})\,,
\end{equation}
where $D[\mathbf{s}_{A},\mathbf{s}'_{A}]$ is the Hamming distance between $L_{\text{qubits}}$-long bitstrings $\mathbf{s}_{A}$ and $\mathbf{s}'_{A}$, i. e., $D[\mathbf{s}_{A},\mathbf{s}'_{A}]\equiv \#\{i\in A|s_i\neq s'_i\}$. Consequently the ensemble average of $X_{a}$, denoted by $\overline{X}$,
\begin{equation}
    \overline{X} = \frac{1}{N_{U}}\sum_{a=1}^{N_{U}}X_{a}\,,
    \label{eq:RM-estimated-purity}
\end{equation}
is the second-order cross-correlations across the ensemble of discrete $N_{U}$ random unitaries $\hat{U}_{a}$, and provides the estimation of the purity $\mathrm{Tr}(\rho_{A}^{2})$ associated with the subsystem A. Finally, the R\'enyi entropy is 
\begin{equation}
    S^{(2)}_{A} = -\mathrm{log}\overline{X}
    \label{eq:renyi-RMP}
\end{equation}

\subsection{Experimental Results}
\label{sec:EEM_results}

We focus on the two-flavor lattice Gross-Neveu model with system size $N_{\text{qubits}}=20$ and the left half-length subsystem of it as the subsystem A with $L_{\text{qubits}}=10$. Now, a primary challenge of the RM protocol is its reliance on an ensemble of local random unitaries to reduce the variance of entropy estimation. This requirement significantly increases the total circuit count when combined with error mitigation techniques such as Pauli Twirling (PT) and Zero-Noise Extrapolation (ZNE). In our implementation on IBM devices, we utilize $60$ distinct RM circuit instances. Incorporating ZNE (using three scale factors: $1, 3,$ and $5$) and PT (with $10$ pair-sampling instances) multiplies the workload accordingly. Consequently, the experiment requires a total of $1,800$ circuits ($60 \times 3 \times 10$) for each entanglement measurement.

Therefore, to maximize throughput on the IBMQ system, we utilize Quantum Multi-Programming (QMP) for parallel circuit execution, which allows for the concurrent processing of multiple quantum circuits, regardless of their specific architecture or depth. This technique has proven effective across diverse quantum applications, such as Grover’s search, quantum amplitude estimation, the quantum support vector machine, and R\'enyi entropy measurements \cite{park2023quantum, rao2024quantum, baker2024parallel, chowdhury2025capturing, chowdhury2025probing, choi2026quantum}.
While QMP enhances the execution efficiency, it introduces challenges such as inter-circuit crosstalk. To mitigate these effects, we follow the strategy proposed in Ref.~\cite{ohkura2022simultaneous, park2023quantum}, where we parallelize two circuits per QMP package, separated by a single idle physical qubit to suppress crosstalk on IBM quantum hardware.
Consequently, the complete set of 1,800 circuits is executed via 900 parallelized pairs. Upon completion of the QMP execution, both circuits in each pair are measured simultaneously. We then post-process the resulting measurement outcomes according to the methodology detailed in Section C in the Supplemental Material of Ref.~\cite{baker2024parallel}.
\begin{figure}[h!]
    \centering
    \includegraphics[width=0.6\linewidth]{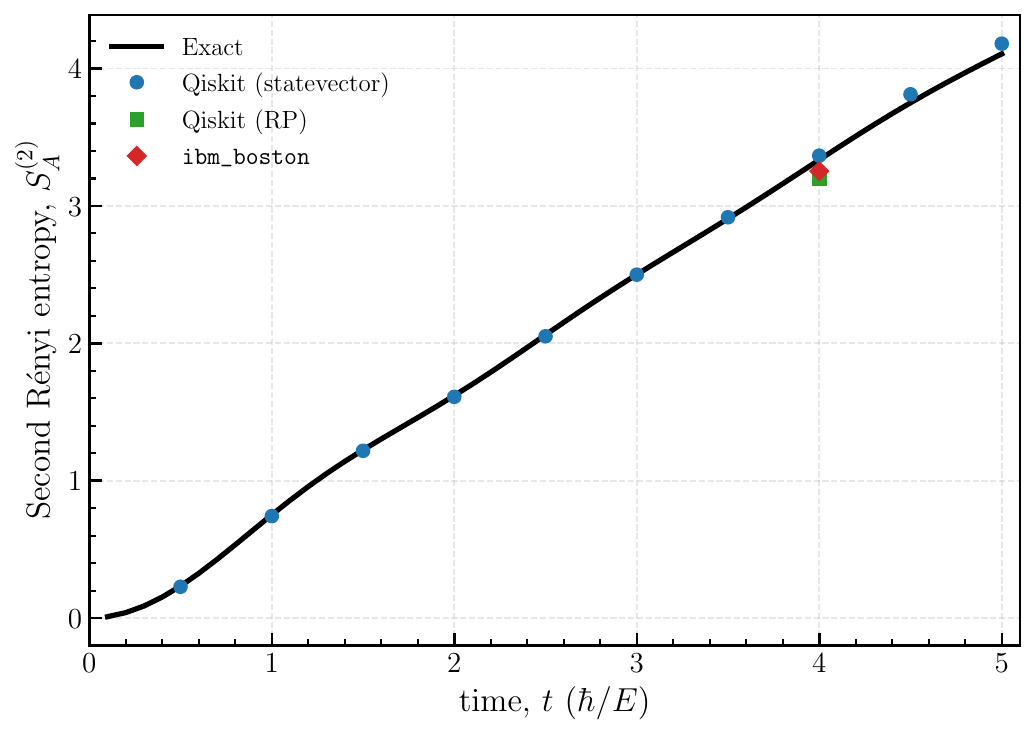}
\caption{The second R\'enyi entropy, $S^{(2)}_{A}$ of the subsystem A (left half-chain) with $L_{\text{qubits}}=10$ in two-flavor lattice Gross-Neveu system with total $N_{\text{qubits}}=20$. Here, the exact computation (black solid line) and Qiskit statevector simulation using optimized second-order Trotterization circuits (blue markers) agree well. At $t=4$, we see further agreement with the noiseless Qiskit simulation (green marker) and the real \texttt{ibm$\_$boston} (red marker) for the RM protocol, with the exact and statevector simulation results.}
\label{fig:S2_comparison}
\end{figure}

The entanglement entropy measurement results are presented in Fig. \ref{fig:S2_comparison}, where we can see that the dynamics of the second R\'enyi entropy $S^{(2)}_{A}$ of the subsystem A computed using Qiskit Statevector simulation with the optimized second-order Trotterization circuits given in Eq.~(\ref{eq:second-Trotter-3}) agrees well with the exact values. As the large ensemble of circuit executions, even with the QMP, requires substantial QPU resources, we focus only on time $t=4$. At $t = 4.0$, the $S^{(2)}_{A}$ values from the exact computation, Qiskit Statevector simulation, noiseless Qiskit RM Protocol simulation, and \texttt{ibm\_boston} are $3.3359$, $3.3659$, $3.1989$, and $3.2533$, respectively. We use the same circuit ensemble from the noiseless Qiskit simulation of the RM protocol in \texttt{ibm\_boston}. Here, we see that the Qiskit Statevector simulation exhibits a $0.90\%$ relative error compared to the exact baseline. In contrast, the Qiskit RM Protocol shows a $4.96\%$ relative error compared to the Statevector result. Notably, the \texttt{ibm\_boston} hardware data deviates by only $1.70\%$ from the Qiskit RM Protocol simulation after mitigating hardware-induced noise using the techniques described in Appendix \ref{app:error_mitigations}. These findings indicate that the entanglement entropy measured on the quantum hardware is in close agreement with noiseless classical benchmarks.

\section{Conclusion and Outlook}\label{sec:conclusion}

In this work, we have presented a comprehensive framework for simulating the real-time dynamics of the multi-flavor Gross–Neveu model on superconducting quantum computers at the utility scale. By combining scalable Trotterization strategies with hardware-aware circuit design, we demonstrated that it is possible to efficiently implement time evolution of an interacting quantum field theory on systems exceeding one hundred qubits. Our construction ensures that the circuit depth per Trotter step scales primarily with the number of fermion flavors rather than the total system size, making it well-suited for near-term quantum devices with limited qubit connectivity.

A central contribution of this work is the development of the Localized Diagonal Operator Approximation (LDOA), which significantly reduces the overhead associated with long-range quartic interactions. By reformulating diagonal unitary synthesis as a structured least-squares problem, we derived compact diagonal circuits that approximate the target quartic circuit with high fidelity while substantially lowering circuit depth and two-qubit gate counts. This improvement is critical for mitigating decoherence and enhancing the feasibility of large-scale simulations on current quantum hardware having limited qubit connectivity. Importantly, the LDOA framework is not restricted to the specific Gross-Neveu model considered here, but is generally applicable to diagonal quantum operators with long-distance interactions. Its practical relevance is particularly significant for quantum hardware with limited connectivity, such as superconducting architectures, where nonlocal interactions must be compiled into local gate sets. In contrast, for all-to-all connected architectures, such approximations are not necessary.

We validated our approach through implementations on IBM superconducting quantum processors, benchmarking the results against classical methods including exact diagonalization and tensor network simulations. The observed agreement in key physical observables, such as the density-density correlation, confirms the reliability of our method in capturing nontrivial real-time dynamics. Furthermore, our simulations provide insight into quantum information spreading and entanglement growth in the Gross–Neveu model, highlighting the potential of quantum computers to explore regimes inaccessible to classical techniques.

Overall, this work establishes a scalable and experimentally validated pathway toward quantum simulation of interacting fermionic quantum field theories. Our results underscore the growing capability of near-term quantum devices to address classically intractable problems and pave the way for future studies of dynamical symmetry breaking, nonequilibrium phenomena, and strongly correlated systems in high-energy and condensed matter physics. Additionally, extending this framework to more complex quantum field theories, higher dimensions, and gauge theories represents an important direction for future research.

\section*{Acknowledgment}
\noindent
K.Y. is supported by the Brookhaven National Laboratory LDRD 24-061, 25-033, and the U.S. Department of Energy, Office of Science, Grants No. DE-SC0012704. K.K. is supported in part by US DOE DE-SC0024673. 
This research used quantum computing resources of the Oak Ridge Leadership Computing Facility, which is a DOE Office of Science User Facility supported under Contract DE-AC05-00OR22725.
This research used resources of the National Energy Research Scientific Computing Center (NERSC), a Department of Energy Office of Science User Facility under Contract No. DE-AC02-05CH11231 using NERSC award DDR-ERCAP0037352 and DDR-ERCAP0037323.

\appendix

\section{Quantum Error Mitigation}\label{app:error_mitigations}

A significant bottleneck in executing quantum algorithms on contemporary hardware, such as IBM Quantum processors, is the ubiquity of noise and operational errors. While quantum error correction (QEC) has been developed to address these limitations, it remains impractical for large-scale problems on current noisy quantum devices. Implementing QEC requires a substantial qubit overhead that exceeds the capacity of modern systems, even when considering recent optimizations \cite{kivlichan2020improved, lee2021even}.

Conversely, quantum error mitigation (QEM) operates within the constraints of noisy, near-term devices by utilizing pre- and post-processing or circuit-level modifications to minimize noise-induced bias. Unlike the heavy overhead associated with QEC, QEM is highly compatible with current hardware due to its minimal qubit requirements. A wide array of QEM strategies—from zero-noise extrapolation to symmetry-based verification—has been successfully validated in recent experimental benchmarks \cite{Kim-error-mitigation, kim2023evidence, charles2305simulating, chowdhury2024enhancing, choi2026quantum, chowdhury2026quantum}. To mitigate the influence of inherent hardware imperfections in our experiments, we employ an integrated suite of four complementary techniques: Twirled Readout Error Extinction (TREX) for measurement error mitigation, Dynamical Decoupling (DD) to suppress decoherence during idle times, Pauli Twirling (PT) to stochasticize coherent errors, and Zero-Noise Extrapolation (ZNE) to approximate noiseless expectation values. Each methodology is detailed in the subsequent sections.


Twirled Readout Error Extinction (TREX) provides a robust framework for mitigating measurement errors without requiring a detailed noise model \cite{PhysRevA.105.032620}. The protocol involves applying random Pauli-X gates to selected qubits immediately prior to measurement, with a corresponding bit-flip applied to the classical output. While redundant in an ideal noise-free environment, this process "twirls" complex, state-dependent readout noise into a uniform, stochastic scaling factor. This simplification effectively decouples individual qubit errors, making them significantly easier to correct. Although TREX is less effective at mitigating strongly correlated multi-qubit errors, its model-independent nature makes it a highly resilient solution for modern hardware. For our experiments, we utilize 10 samples for the TREX process.


Dynamical Decoupling (DD) is a well-established technique for suppressing decoherence and crosstalk, particularly noise arising from interactions with spectator qubits. By applying periodic sequences of high-frequency control pulses, DD averages out system-environment couplings, effectively neutralizing their impact on the system's error dynamics \cite{viola1999dynamical}. In practice, this involves applying single-qubit rotations to idle qubits to toggle their basis, thereby extending the effective coherence time of the circuit. The utility of DD has been extensively validated across various experimental platforms \cite{ezzell2022dynamical, niu2022effects, Kim-error-mitigation, kim2023evidence, charles2305simulating}. In our implementation, we insert a $(t/4, X, t/2, X, t/4)$ sequence into every idling period, where $X$ denotes an $\texttt{XGate}$ and $t$ represents the total idling duration excluding the execution time of the two $\texttt{XGate}$ pulses.


Pauli Twirling (PT) is a technique designed to transform structured, coherent errors into stochastic Pauli noise. By scrambling coherent interactions through an ensemble of operations, PT eliminates the off-diagonal elements of the error channel in the Pauli basis $\{ I, \sigma^x, \sigma^y, \sigma^z \}$, resulting in a simplified noise model that is more amenable to mitigation \cite{bennett1996purification, wallman2016noise, cai2019constructing}. In practice, PT involves conjugating a Clifford gate with Pauli operators such that the overall operation remains algebraically equivalent to the original gate. This approach has been successfully validated in several recent studies \cite{Kim-error-mitigation, kim2023evidence, chowdhury2024enhancing, chowdhury2025capturing, chowdhury2025probing}. We specifically apply PT to the CZ gates in our circuits. To do so, we consider the 256 possible combinations ($4^4$) of Pauli gates surrounding the two-qubit gate. For each implementation, we generate ten distinct circuit instances, each utilizing a randomly sampled Pauli-twirling sequence from the set of mathematically equivalent configurations.


Zero-Noise Extrapolation (ZNE) is an error mitigation strategy that approximates noise-free expectation values by extrapolating data from measurements taken at systematically scaled noise levels \cite{Temme-error-mitigation, Li-error-mitigation, giurgica2020digital}. In our implementation, we utilize local unitary gate folding \cite{giurgica2020digital} specifically for two-qubit CZ gates, as their error rates are empirically an order of magnitude higher than those of single-qubit gates. We achieve noise amplification by folding the CZ operations with scaling factors of 1 (noiseless baseline), 3 (single fold), and 5 (double fold), thereby increasing the effective noise while preserving the circuit's logical identity. While ZNE assumes that noise can be controlled and amplified linearly or polynomially, real-world quantum noise is often non-Markovian or time-dependent. Such instabilities can lead to unpredictable circuit outputs, potentially rendering the extrapolation unreliable. Furthermore, while fitting curves derived from small-scale systems can theoretically be extended to larger systems, this assumption of noise-model consistency often breaks down as the qubit count increases.

\bibliography{references}
\end{document}